\documentclass[%
 reprint,
superscriptaddress,
 amsmath,amssymb,
 aps,prx
]{revtex4-2}

\usepackage{graphicx}% Include figure files
\usepackage{dcolumn}% Align table columns on decimal point
\usepackage{bm}% bold math
\usepackage[colorlinks=true,linkcolor=blue,urlcolor=blue,citecolor=blue]{hyperref}% add hypertext capabilities

\usepackage{subfigure}
\usepackage{xcolor}
\usepackage{amsmath}
\usepackage{physics}
\usepackage{float}

\usepackage{graphicx}% Include figure files
\usepackage{dcolumn}%
 \graphicspath{ {./Images/} }
\begin{document}

\preprint{APS/123-QED}
\title{Compass state: Effect of squeezing and displacement on the Fock space}

\author{Arman}
\email{a19rs004@iiserkol.ac.in}
\affiliation{Department of Physical Sciences, Indian Institute of Science Education and Research Kolkata, Mohanpur-741246, West Bengal, India}

\author{Prasanta K. Panigrahi}%
\email{pprasanta@iiserkol.ac.in}
\affiliation{Department of Physical Sciences, Indian Institute of Science Education and Research Kolkata, Mohanpur-741246, West Bengal, India}

\date{\today}

%is prone to decoherence effect as compared to the prosed single mode states . terms: squeezing and displacement.

\date{\today}

\begin{abstract}
We investigate a broad class of non-classical states, composed of superposed squeezed and displaced number states. The phase space structure is analysed, keeping in mind, Heisenberg limited sensitivity in parameter estimation. Appropriate squeezing and displacement parameters are identified, wherein state fidelity in comparison to metrologically sensitive compass state, is more than 99$\%$. Also, the variance in small shifts measurements, is found equal for both proposed and compass states. They show similar behaviour with change in average photon number. In metrological application, number variance of the compass state being small for low coherent amplitude, suggests its potential to estimate damping parameter. 

\end{abstract}

\maketitle
\section{Introduction}
Coherent states are minimum uncertainty state with Poissonian statistics in the Fock space, describe their classical nature. The superposition of four different coherent states \cite{Zurek2001}, is known as the compass or kitten state (KS). Note that in this paper, terminology of compass and kitten state is same. The kitten state possesses sub-Planck scale structures in the phase space with oscillatory distribution as a function of small quadrature phase space shift ($\delta x, \delta p$). The interferometric phase space structure makes the KS state sensitive to quadrature fluctuation, therefore allowing detection of weak external forces \cite{dalvit,pk1} as well as quadrature phase fluctuation \cite{pk2,Hl-zurek}. This behavior of oscillation can be realized in the experiment by measuring the inversion of two level system (TLS) entangled with the cavity system maintained in the compass state as a probe \cite{david-wigner,luis,Hl-zurek}. In this process, atom and field system interact in a way such that inversion of TLS becomes proportional to the Wigner function of the system. This whole process allows to measure the sensitivity  of the probe system in terms of small phase-space perturbation \cite{luis}. Due to high sensitivity of KS as loss of the interferometric structure or non-classicality in the presence of damping to the environment, it presents great potential as a probe in estimating damping constant of the system. %.Use a probe in quantum metrology to estimate the unknown parameter i.e. damping constant for the open systems causing loss of non-classicality (the Wigner negativity) for this state with damping constant is prominent. 

Need for generation of highly non-classical states \cite{pac-agarwal,pac-arman,biagi2022photon} for advancement in fault-tolerant continuous variable  quantum computation, information processing and communication, have become pertinent in the optical as well as microwave \cite{puri2017engineering} 
platform. The cat, compass and higher order of $n$ superposition coherent states (Quantum hypercube state \cite{QHS}) have found their uses as cat codes in quantum error correction (QEC) due to the property of returning back to same state after two, four and $n$ excitation loss,  respectively \cite{jiang-code,reviw-cai,quant-code}. Encoding logical qubit in higher dimensional Hilbert space of bosonic systems, suffering a few natural errors, has allowed to prepare QEC systems, preserving them over longer period of time and achieving a fault-tolerant systems. Large amplitude of these bosonic states increases the success of correcting errors in the logical qubit. There have been many proposal \cite{biswas,choudhury2011proposal,tarachaturbedi} and experimental \cite{ex-cat-nelson} generation  of low amplitude cat states in microwave \cite{kirchmair2013observation} as well as optical \cite{ourjoumtsev2007generation} platform, however, with very few proposal \cite{mikheev2019efficient, hastrup}, it remains a challenge for the preparation of high fidelity large amplitude cat, KS as well as $n$-coherent state. This motivates us to find new methods for producing state with an achievable and possibly optical setup.     

In this paper, we study the phase-space structures (Sec.\hyperref[sec:III]{III}) for two states that are proposed in Sec.\hyperref[sec:II]{II}. The probability distribution is studied for both squeezing and displacement parameter of state. The squeezing is not limited to its uses in single mode as is shown recently for two-mode squeezing superposition \cite{twomode} with metrological aspect and their ion trap implementation. Further, with the help of fidelity between proposed states and compass state, we investigate their similarities and sensitivity in the Sec.\hyperref[sec:IV]{IV} and hence, theoretical model for their preparations  in Sec.\hyperref[sec:V]{V}. Finally, Sec.\hyperref[sec:VII]{VII} contains conclusion with summary.

\section{Superposition of squeezed and displaced number states}\label{sec:II}

% We study properties of squeezed states involving initial values number states. Different kind of non-classical properties are calculated to find resemblance with kitten or compass states.  Two kind of different states have been proposed to find similarities in phase space structure of compass state. 
Main aim for this paper is finding the form of states in such a way that there exists a large overlap of KS with the states comprised of squeezing, displacement and Fock number parameters. We know that the photon subtracted squeezed state ($\hat{a}S[r]\ket{0}$) is equivalent to the squeezed single photon state ($S[r]\ket{1}$), describes two  Gaussian lobes with negative center in the phase space causing large overlap with odd cat state \cite{biswas}. Using this analogy, we have two ways to obtain KS with possibly large overlap: either (i) application of squeezing on two different displacement for Fock state or (ii) two opposite phase-space oriented squeezing of the Fock state. 
Before investigating qualitative properties - the probability number distribution and  the phase-space distribution, and  quantitative ones - fidelity, we write the form of two proposed states in the $x$-projected space. First state involves combination of squeeze and displacement operators acting on the number state. Thus, squeezed superposed displaced number state (SSDNS) with displacement (`$\alpha$') and squeezing (`$r$') parameters, is written  in the form
$$\ket{\psi_{1}}=N_{1}S[r](D[\alpha]+D[-\alpha])\ket{n},$$ 
where  $N_{1}$ and $D[\alpha]$ are the normalization and the displacement operator with $\alpha$ being real, respectively. 
Other state is the superposition of two squeezed number states with opposite signs of squeezing parameter $r$. We write superposed squeezed number state (SSNS) as $$\ket{\psi_{2}}=N_{2}(S[r]+S[-r])\ket{n},$$
where  $N_{2}$ and $S[r]$ are the normalization and the squeezing operator of the form  $e^{r\frac{a^{2}-{a^{+}}^{2}}{2}}$ with $r >0,$ respectively. Knowing the form of normalized oscillator state in terms of Hermite polynomial $H_{n}(x)$ of order $n$, in $x$-projection,
$$\phi_{n}(e^{r}x-\alpha)=\frac{\sqrt{e^{r}}e^{-\frac{1}{2} \left(-\sqrt{2} \alpha +e^r x\right)^2} H_n\left(e^r x-\alpha  \sqrt{2}\right)}{\sqrt{2^{n}n!\sqrt{\pi}}};$$
we write SSDNS with real displacement parameter $
`\alpha'$
$$\psi_{1}(x)=\bra{x}\ket{\psi_{1}}=N_{1}\left(\phi_{n}(e^{r}x-\alpha)+\phi_{n}(e^{r}x+\alpha)\right),$$ and SSNS depending on 
squeezing parameters ($\pm r$)
$$\psi_{2}(x)=\bra{x}\ket{\psi_{2}}=N_{2}\left(\phi_{n}(e^{r}x)+\phi_{n}(e^{-r}x)\right).$$

The normalization $N_{1}$ and $N_{2}$ for both states are given as
$$\begin{aligned}N_{1}^{-2}=\bar{I}_{n}(r,\alpha,-\alpha)+\bar{I}_{n}(r,-\alpha,\alpha)+\bar{I}_{n}(r,-\alpha,-\alpha)\\+\bar{I}_{n}(r,\alpha,\alpha)\end{aligned}$$
and $$\begin{aligned}
N_{2}^{-2}=I_{nn}(r,-r,0,0)+I_{nn}(r,-r,0,0)+\bar{I}_{n}(r,0,0)\\+\bar{I}_{n}(-r,0,0),\end{aligned}$$
 
where symbols, for $r\neq \bar{r},$ $$I_{nm}(r,\bar{r},\alpha,\beta)=\int_{-\infty}^{\infty}dx\, \phi_{n}(e^{r}x-\sqrt{2}\alpha)\phi_{m}^{*}(e^{\bar{r}}x-\sqrt{2}\beta)$$
and for $r= \bar{r},$$$\bar{I}_{n}(r,\alpha,\beta)=\int_{-\infty}^{\infty}dx \phi_{n}(e^{r}x-\sqrt{2}\alpha)\phi_{n}^{*}(e^{r}x-\sqrt{2}\beta).$$
The above integrations are evaluated using identity, obtained through generating function of the Hermite polynomials \cite{2dhermite}, as follows (see \hyperref[appendix:a]{Appendix}):
$$\int_{-\infty }^{\infty } H_n(A x) H_m(B x-d) \frac{e^{ \left(-\frac{1}{2} (A x)^2+\delta  (B x-d)-\frac{1}{2} (B x-d)^2\right) }}{e^{-\gamma  (A x)}}dx $$
\begin{multline}\label{identity1}
    = \sqrt{\left(\frac{A^2-B^2}{A^2+B^2}\right)^{m} \left(\frac{B^2-A^2}{A^2+B^2}\right)^{n}} \frac{\sqrt{2 \pi}e^{ -\frac{(B \gamma -A (d+\delta ))^2}{2(A^2+B^2)}}}{\sqrt{A^2+B^2}e^{-\frac{\gamma ^2+\delta ^2}{2}}}
\\  \underset{j=0}{\overset{\min (m,n)}{\sum }}\frac{m! n! (4 AB)^{j}\,H_{m-j}(X_{1})H_{n-j}(X_{2})}{j! (m-j)! (n-j)!\left(\sqrt{-\left(A^2-B^2\right)^2}\right)^j},
\end{multline}
where $A$ and $B$ constants belong to real domain, and quantities
$X_{1}=
\left(\frac{B \gamma  A-A^2 d+B^2 \delta }{\sqrt{A^4-B^4}}\right),\,X_{2}= \left(\frac{A (A \gamma +B (d+\delta ))}{\sqrt{B^4-A^4}}\right).
$
When $A=\pm B$, the identity in Eq.\ref{identity1}  becomes
$$\int_{-\infty }^{\infty } H_n(A x) H_m(\pm A x-d) \frac{e^{ \left(\delta  (\pm A x-d)-\frac{1}{2} (\pm A x-d)^2\right) }}{e^{\frac{1}{2} (A x)^2-\gamma  (A x)}}dx $$
\begin{multline}\label{identity2}
    =\frac{\sqrt{ \pi}}{\abs{A}} \frac{e^{ -\frac{( \pm\gamma -d-\delta )^2}{4}}}{e^{-\frac{\gamma ^2+\delta ^2}{2}}} \underset{j=0}{\overset{\min (m,n)}{\sum }}\frac{m! n! 2^{j}\left(\pm\gamma- d+\delta \right)^{m-j}}{j! (m-j)! (n-j)!}
\\ \times\frac{\left(\pm \gamma + d+\delta \right)^{n-j}}{j! (m-j)! (n-j)!}.
\end{multline}

\begin{center}
    \begin{figure}[H]
    \begin{minipage}[b]{0.47\linewidth}
    \centering
    \includegraphics[width=1.07\textwidth,height=.9\textwidth]{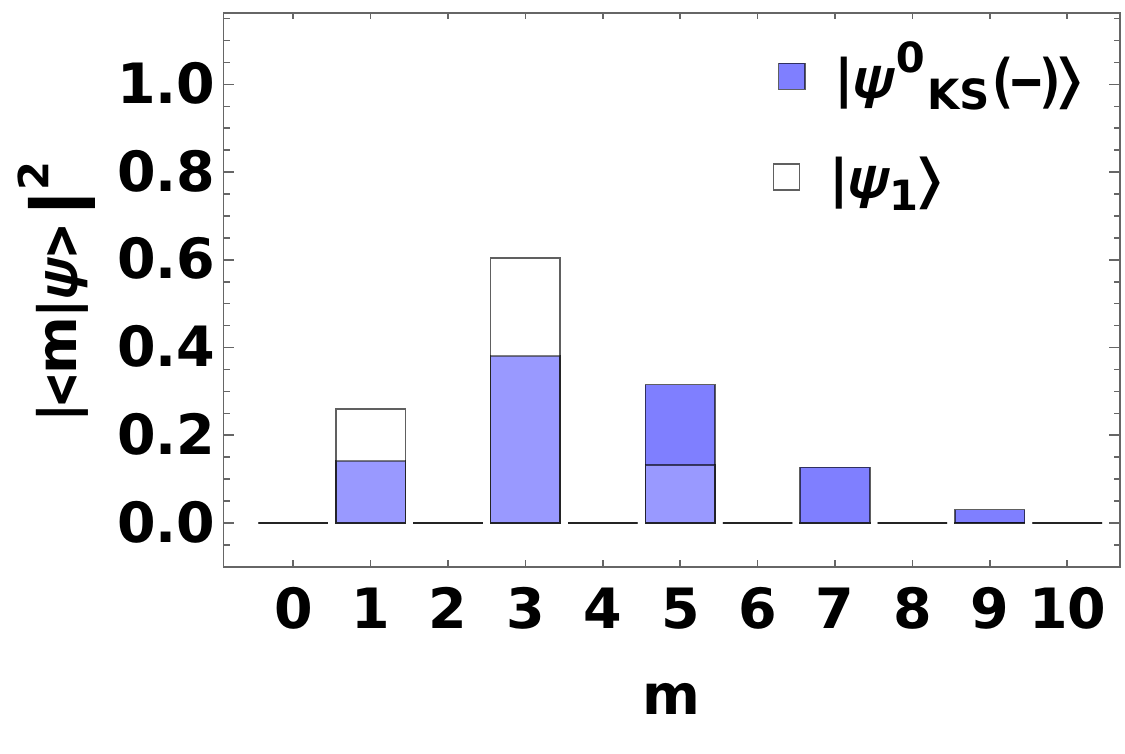}
    \end{minipage}
    \hspace{0.05cm}
    \begin{minipage}[b]{0.47\linewidth}
    \centering
     \includegraphics[width=1.07\textwidth,height=.9\textwidth]{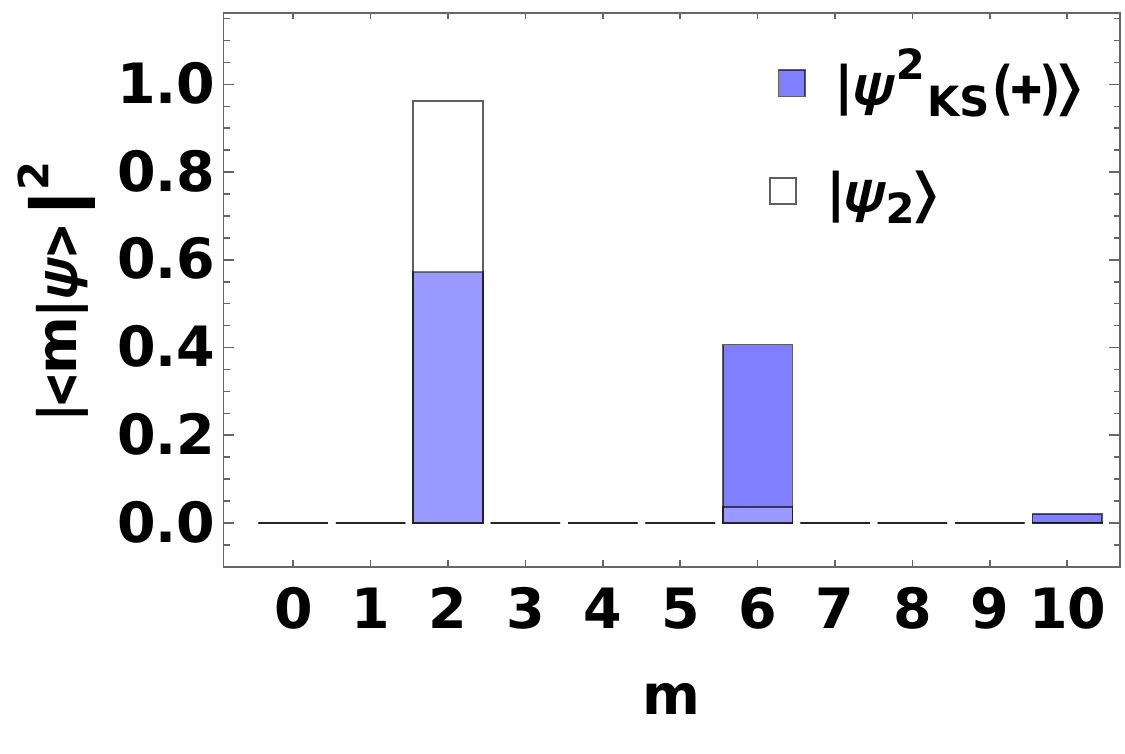}
    \end{minipage}
    \caption{(Color online) The comparison of number distribution against  Fock number $m$ is between SSDNS $\psi_{1}$ with parameters $(r,\alpha,n)=(0.3,1.8,1)$  and KS $\psi^{0}_{KS}(-)$ for coherent amplitude $|\beta|=2$ on the left, while distribution on the right is between SSNS $\psi_{2}$ with parameters $(r,n)=(0.3,2)$  and KS $\psi^{2}_{KS}(+)$.}
\label{numdist}
\end{figure}
\end{center}

First, we expand both of the states in Fock space representation:
$$\ket{\psi_{1}}=\sum_{k=0}^{\infty}c_{k}\ket{k};\,\ket{\psi_{2}}=\sum_{k=0}^{\infty}b_{k}\ket{k},$$
where probability amplitudes are
$$c_{k}=\bra{k}\ket{\psi_{1}}=N_{1}\left(I_{nk}(r,0,\alpha,0) +I_{nk}(r,0,-\alpha,0)\right)$$
and
$$b_{k}=\bra{k}\ket{\psi_{2}}=N_{2}\left(I_{nk}(r,0,0,0) +I_{nk}(-r,0,0,0)\right).$$
Furthermore, to explore and obtain large similarity via qualitative and quantitative properties for the pairs of proposed and KS states, we look for the relevant form of KSs  corresponding to $`n'$ parameter dependant $\psi_{1}$ and $\psi_{2}$. The $l$ orthogonal states can be constructed for $l$ superposed coherent states with an appropriate weight factor \cite{jiang-code}. In Fock space, four orthogonal KSs with $\beta$ amplitude and label $l\in\{0,1,2,3\}$, are written as
$$\ket{\psi^{l}_{KS
}(\pm)}=\ket{\beta}\pm(-1)^{-l}\ket{-\beta} +(i)^{-l}\ket{i\beta} \pm(-i)^{-l}\ket{-i\beta}  $$
$$= \left(\sum_{p=0}^{p=\infty}\abs{f_{p}^{\pm}}^{2}\right)^{-\frac{1}{2}} \sum_{m=0}^{m=\infty}f_{m}^{\pm}\ket{m},$$
where, $$f_{m}^{+}=\frac{\beta^{4n+l}\delta_{m,4n+l}}{\sqrt{(4n+l)!}},\,f_{m}^{-}=\frac{(1+(-1)^{n}i)\beta^{2n+l+1}\delta_{m,2n+l+1}}{\sqrt{(2n+l+1)!}} .$$
In above expression, $l$ ranges from 0 to 3 and 0 to 1 for $\ket{\psi^{l}_{KS}(+)}$ and $\ket{\psi^{l}_{KS}(-)}$, possessing four and two orthogonal states, respectively.
% $$=\left\{\begin{tabular}{cc}
%  % \frac{
%  \left(\sum_{m=0}^{m=\infty}\abs{f_{m}}^{2}\right)^{-\frac{1}{2}} \sum_{m=0}^{m=\infty}f_{m}\ket{m}
%   %}{\sqrt{\sum_{n=0}^{n=\infty}\frac{\beta^{4n+l}}{(4n+l)!}}} 
%   & f_{m}=\frac{\beta^{4n+l}\delta_{m,4n+l}}{\sqrt{(4n+l)!}} \\
%  \left(\sum_{m=0}^{m=\infty}\abs{f_{m}}^{2}\right)^{-\frac{1}{2}} \sum_{m=0}^{m=\infty}f_{m}\ket{m}
%   & f_{m}=\frac{(1+(-1)^{n}\iota)\beta^{2n+l+1}\delta_{m,2n+l+1}}{\sqrt{(2n+l+1)!}}
% \end{tabular}\right.$$
%$$=\frac{e^{-\frac{\abs{\beta}^{2}}{2}}\sum_{n=0}^{n=\infty}\frac{\beta^{4n+l}}{\sqrt{(4n+l)!}}\ket{4n+l}}{2 \sqrt{2 e^{-\left| \beta \right| ^2} \cos \left(\frac{\pi  l}{2}-\left| \beta \right| ^2\right) \cos (2 \Im(\beta ) \Re(\beta ))+ (-1)^l e^{-2 \left| \beta \right| ^2}+1}}$$
It becomes clear for the choice of two different KSs when the Fock number distribution compared with our proposed states. Plots of probability number distributions for both states against Fock state number `$n\,$' can be seen in Fig.\ref{numdist}. Both compass $\ket{\psi^{l}_{KS}(-)}$ and first state $\ket{\psi_{1}}$ possess odd distribution (Fig.\ref{numdist}) for $n=1$ and $l=0$, while their behavior changes  to even distribution for $n$ even and $l=1$. In first case, both states show their distribution similar to the cat state, where non-zero adjacent state are at a step of 2 along Fock state $\ket{m}$ for either even or odd parity. Bottom plot in Fig.\ref{numdist} describes one case of comparison between compass $\ket{\psi^{l}_{KS}(+)}$ and first state $\ket{\psi_{2}}$ for $n=2$ and $l=2$ possessing even distribution at a step of 4 along Fock state $\ket{m}$. Their behavior also changes to even and odd distribution corresponding to even and odd values of $n$ and $l$. Considerable overlap between compass state and our proposed state for number distribution, is obtained by tuning squeezing, displacement and $n$ parameters for appropriate output of the coherent amplitude $\beta$. Further, analyzing the Wigner distribution allows us to view whether there exist interferometric structures with sub-Planck scales and regular oscillations.  

% \begin{figure}[ht]
% \centering
% \includegraphics[width=.4\textwidth]{Main-tex/Images/fisksofonla.pdf}
% \centering
% \includegraphics[width=.4\textwidth]{Main-tex/Images/2ndksevfonla.pdf}
% \caption{Plot shows the comparison of number distribution against $n$ Fock state between SDNS $\psi_{1}$ with parameters $(r,\alpha,n)=(0.3,1.8,1)$  and Compass state $\psi^{0}_{KS}(-)$ of coherent amplitude $\beta=2$ (Top), while (Bottom) displays between  SSNS $\psi_{2}$ with  parameters $(r,n)=(0.3,2)$  and Compass state $\psi^{2}_{KS}(+)$ of coherent amplitude $\beta=2$.}
% \label{numdist}
% \end{figure}
%%%%%%%%%%%%%%%%%%%%%%%%

%%%%%%%%%%%%%%%%%%%%%%%%

\section{Wigner function}\label{sec:III}
The Wigner function is the quasi-probability distribution in the phase space as one of the measure for non-classicality depending upon its negative volume. The presence of positive and negative part in this distribution signifies the displaced parity of the state. The Wigner function for both states are calculated as follows:
$$W_{1}(x,p)=\frac{1}{\pi} \int_{-\infty}^{\infty}dy \bra{x-y}\ket{\psi_{1}}\bra{\psi_{1}}\ket{x+y}e^{2ipy}$$
\begin{multline}
=N_{1}^{2}\left(w_{n}(r,\alpha,\alpha)+w_{n}(r,-\alpha,-\alpha)+w_{n}(r,\alpha,-\alpha)\right.\\\left.+w_{n}(r,-\alpha,\alpha)\right),\end{multline}
and
$$W_{2}(x,p)=\frac{1}{\pi} \int_{-\infty}^{\infty}dy \bra{x-y}\ket{\psi_{2}}\bra{\psi_{2}}\ket{x+y}e^{2ipy}$$
\begin{multline}
=N_{2}^{2}\left(w_{n}(r,0,0)+w_{n}(-r,0,0)+w_{nn}(r,-r,0,0)\right.\\\left.+w_{nn}(-r,r,0,0)\right)\end{multline}

where $W_{1}(x,p)$ and $W_{2}(x,p)$ are the Wigner functions for states $\ket{\psi_{1}}$ and $\ket{\psi_{2}}$, respectively. The functions used above \begin{multline*}
    w_{nm}(r,\bar{r},\alpha,\beta)= \frac{1}{\pi}\int_{-\infty}^{\infty}dy\,\phi_{n}(e^{r}(x-y)-\sqrt{2}\alpha)\\\times\phi_{m}^{*}(e^{\bar{r}}(x+y)-\sqrt{2}\beta)e^{2ipy}
\end{multline*}    
and
\begin{multline*}
    w_{n}(r,\alpha,\beta)= \frac{1}{\pi}\int_{-\infty}^{\infty}dy\,\phi_{n}(e^{r}(x-y)-\sqrt{2}\alpha)\\\times\phi_{n}^{*}(e^{r}(x+y)-\sqrt{2}\beta)e^{2ipy},
\end{multline*}  
are evaluated with help of identities Eq.\ref{identity1} and Eq.\ref{identity2}. Similarly, Wigner function $(W_{KS})$ for the kitten state $\ket{\psi_{KS}^{l}(\pm)}$ includes 16 terms of Gaussian integrals \cite{Zurek2001}. 

\begin{figure}[ht]
\begin{minipage}[b]{0.45\linewidth}
\centering
\includegraphics[width=1.2\textwidth]{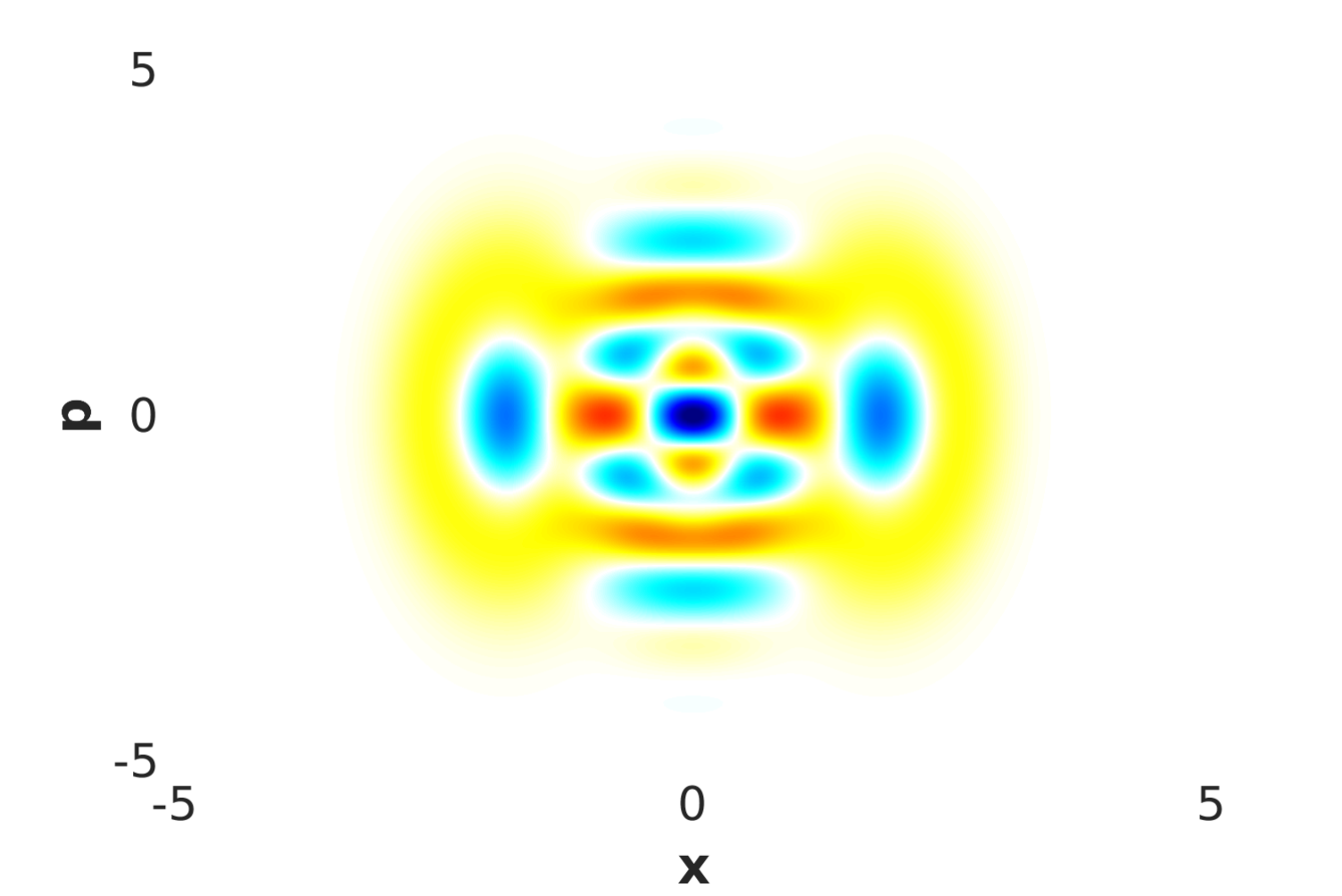}
\end{minipage}
\hspace{0.2cm}
\begin{minipage}[b]{0.45\linewidth}
\centering
\includegraphics[width=1.2\textwidth]{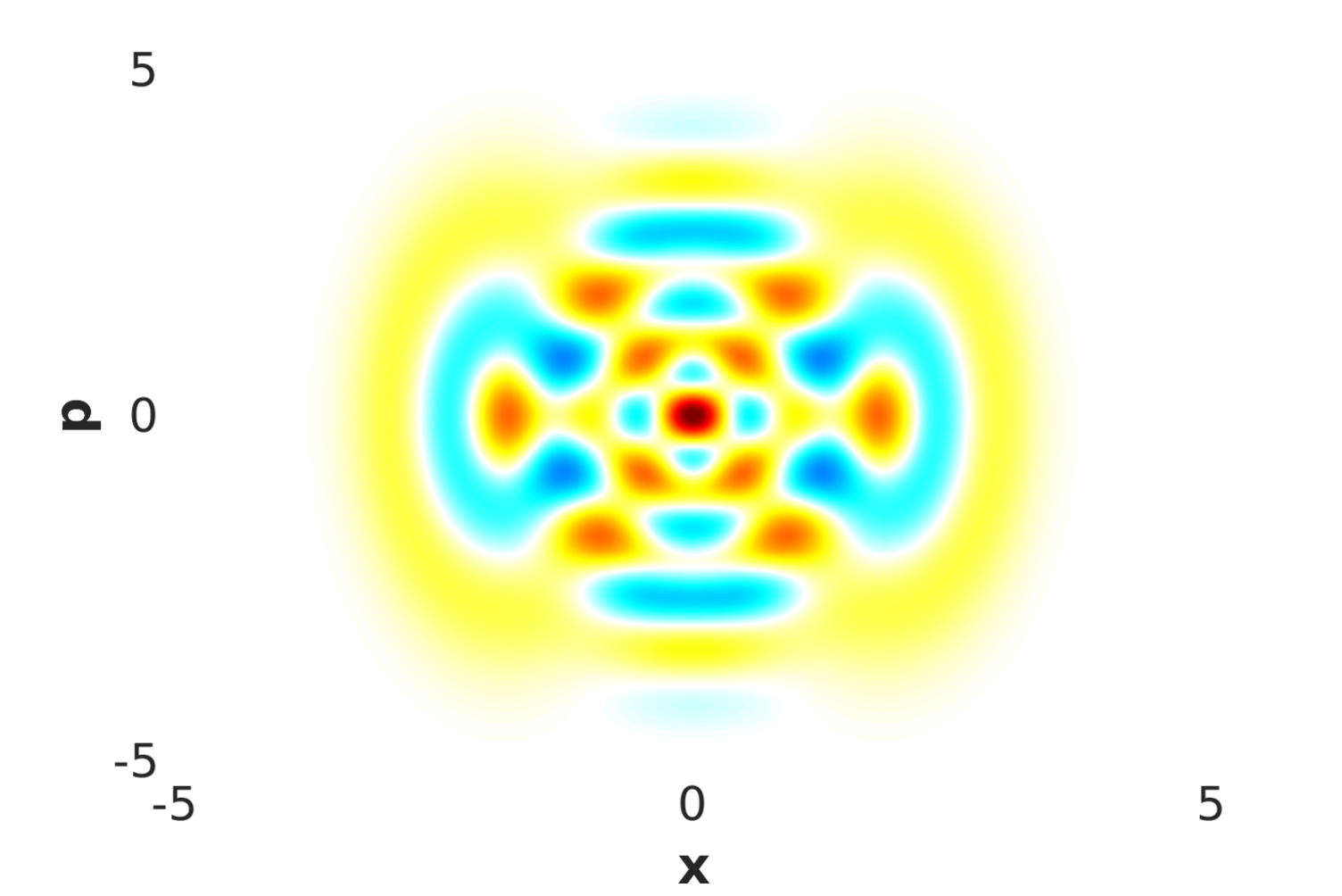}
\end{minipage}
\begin{minipage}[b]{0.45\linewidth}
\centering
\includegraphics[width=1.2\textwidth]{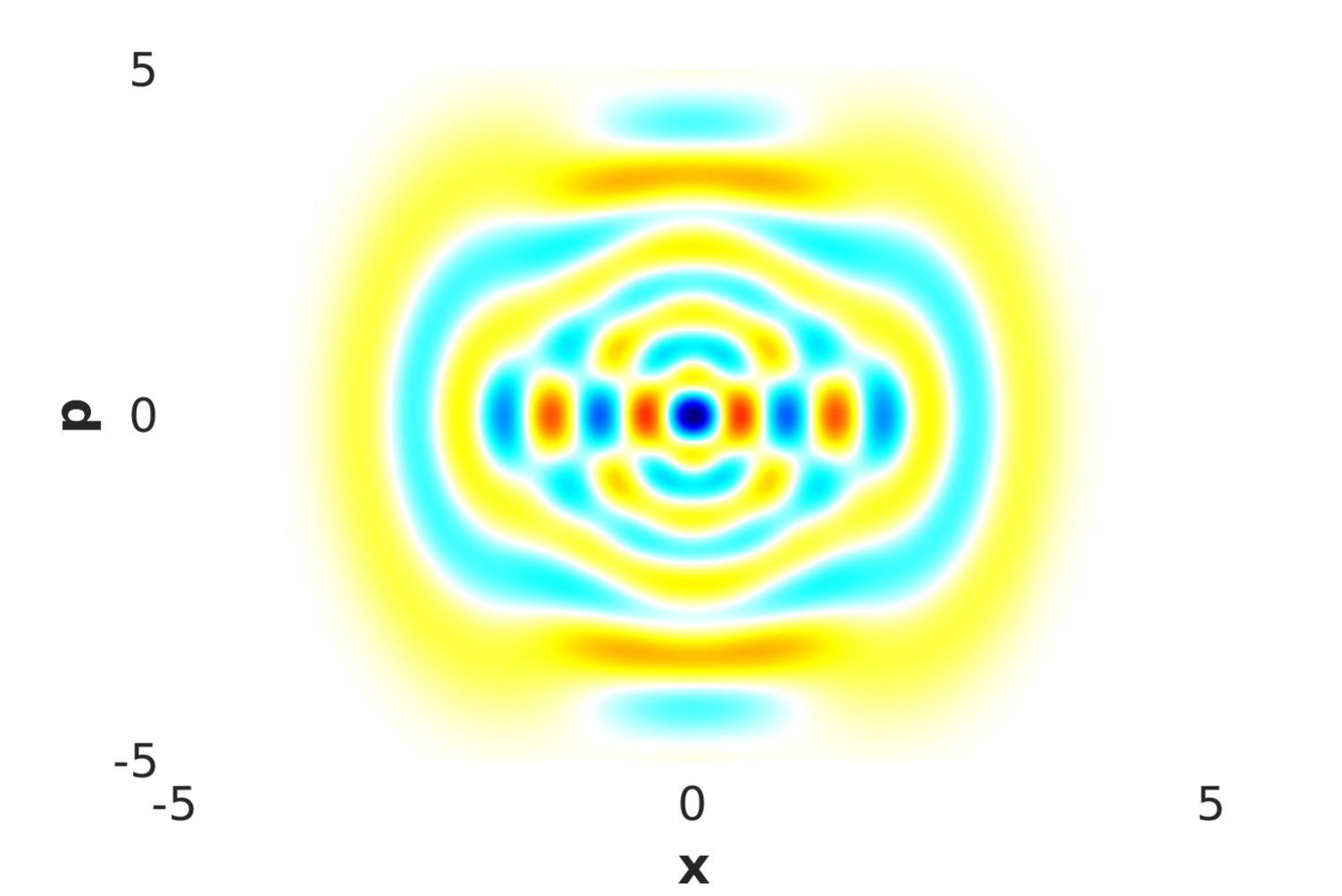}
\end{minipage}
\hspace{0.2cm}
\begin{minipage}[b]{0.45\linewidth}
\centering
\includegraphics[width=1.2\textwidth]{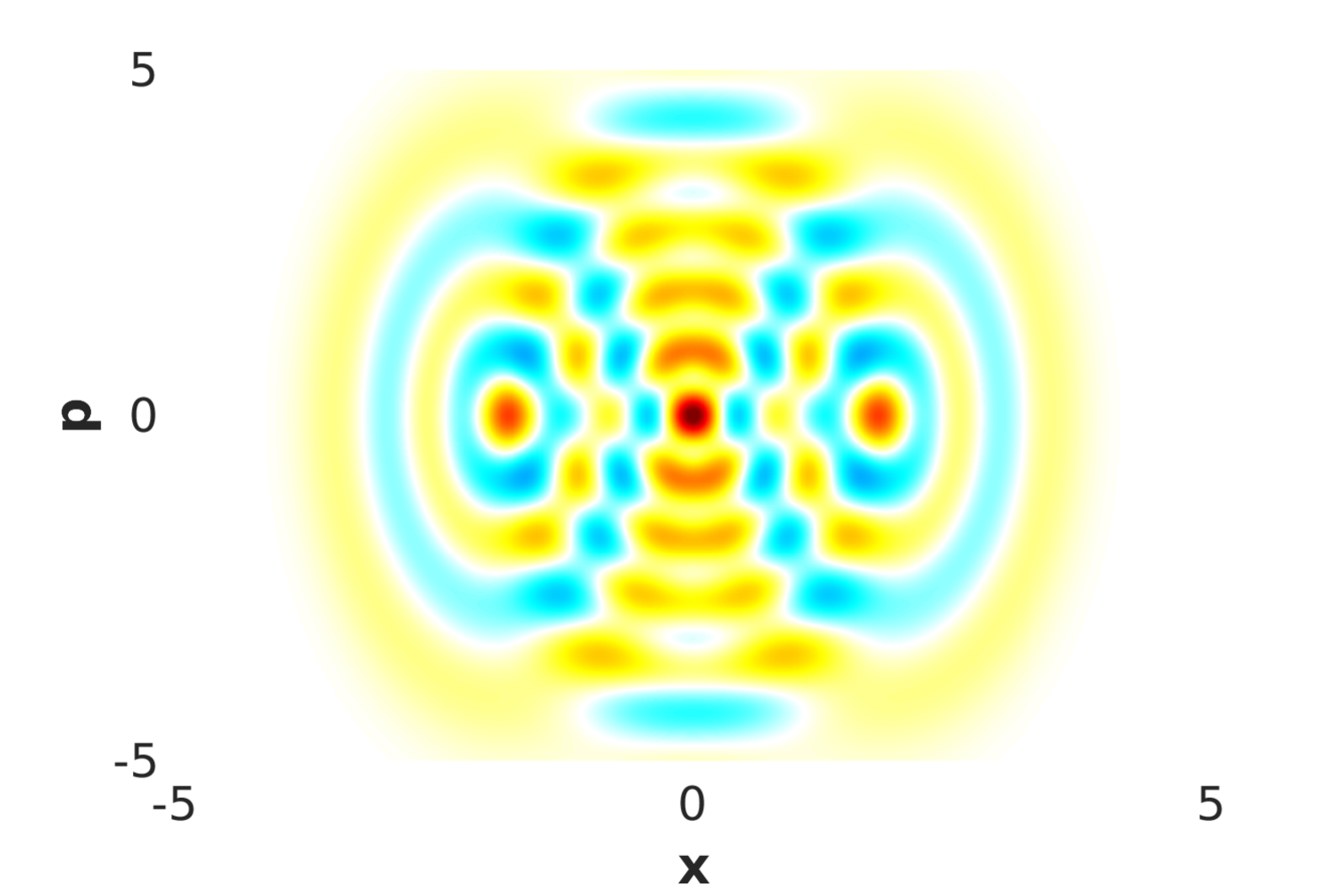}
\end{minipage}
\caption{(Color online) Interference appears in the Wigner function of SSDNS for squeezing parameter $r=0.45$ and  displacement  $\alpha =2$ for different Fock number $n=1$ (Top Left),\, $n=2$ (Top Right),\, $n=3$ (Bottom Left),\,$n=4$ (Bottom Right).}
\label{firststate}
\end{figure}

\begin{figure}[ht]
\begin{minipage}[b]{0.45\linewidth}
\centering
\includegraphics[width=1.2\textwidth]{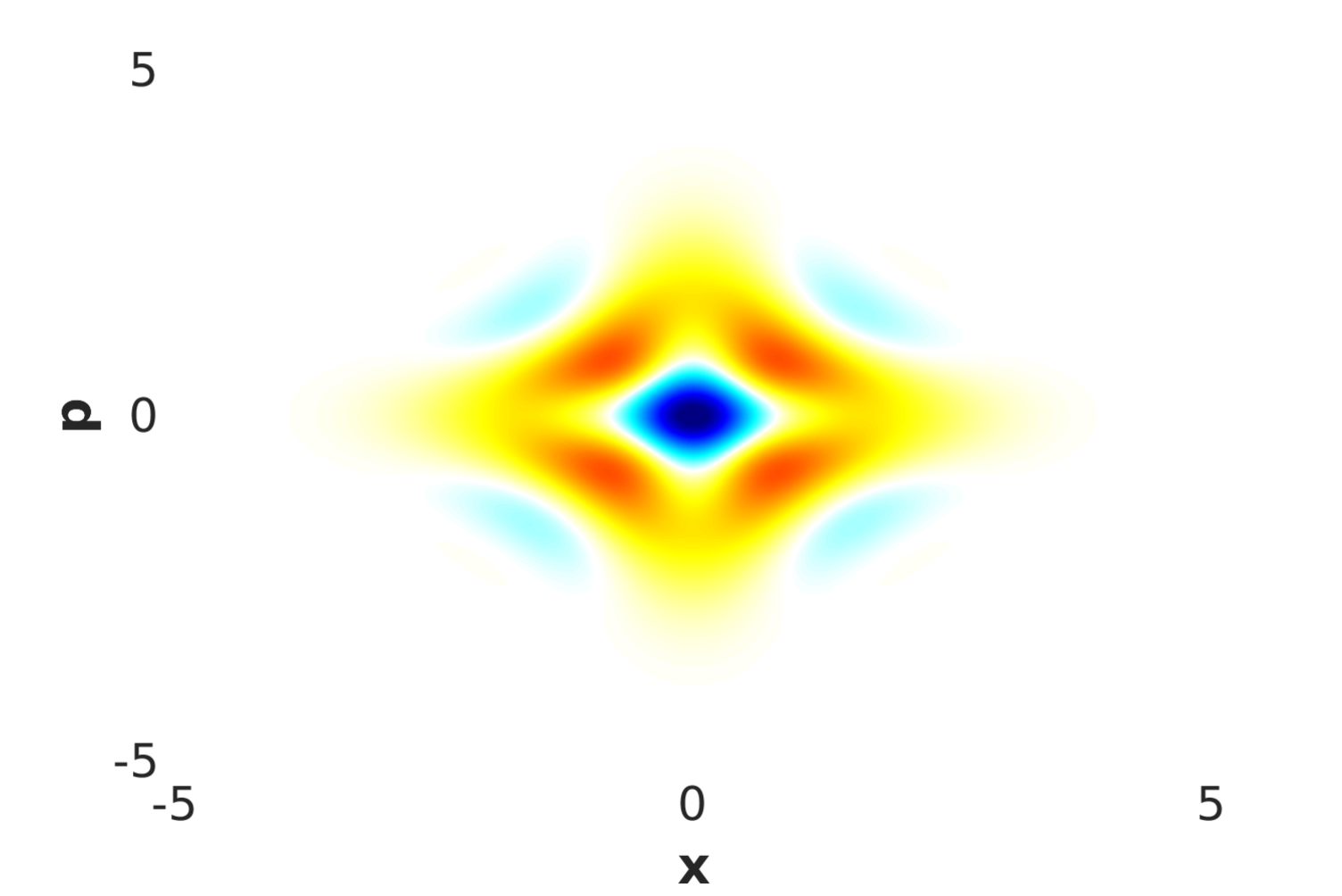}
\end{minipage}
\hspace{0.2cm}
\begin{minipage}[b]{0.45\linewidth}
\centering
\includegraphics[width=1.2\textwidth]{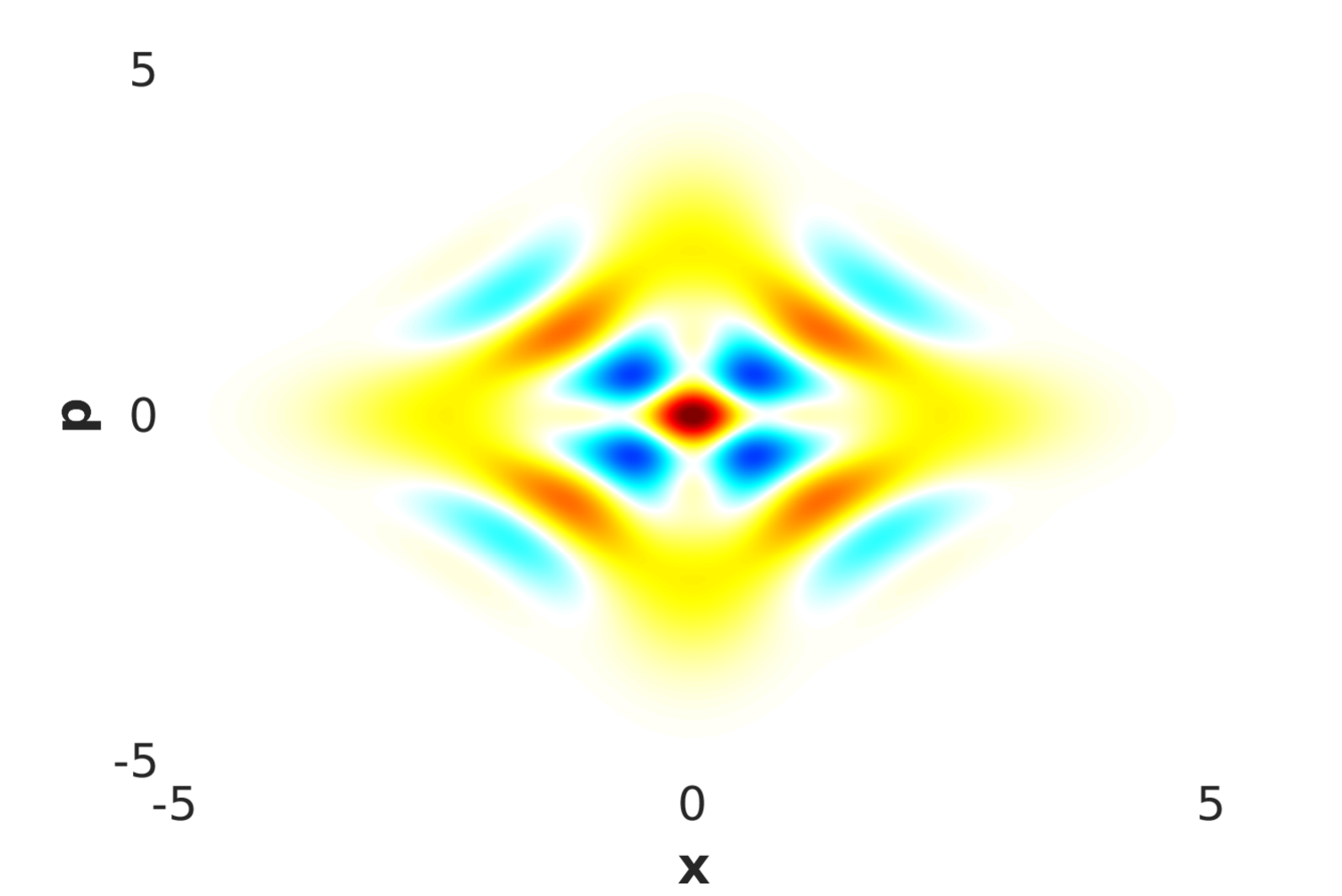}
\end{minipage}
\begin{minipage}[b]{0.45\linewidth}
\centering
\includegraphics[width=1.2\textwidth]{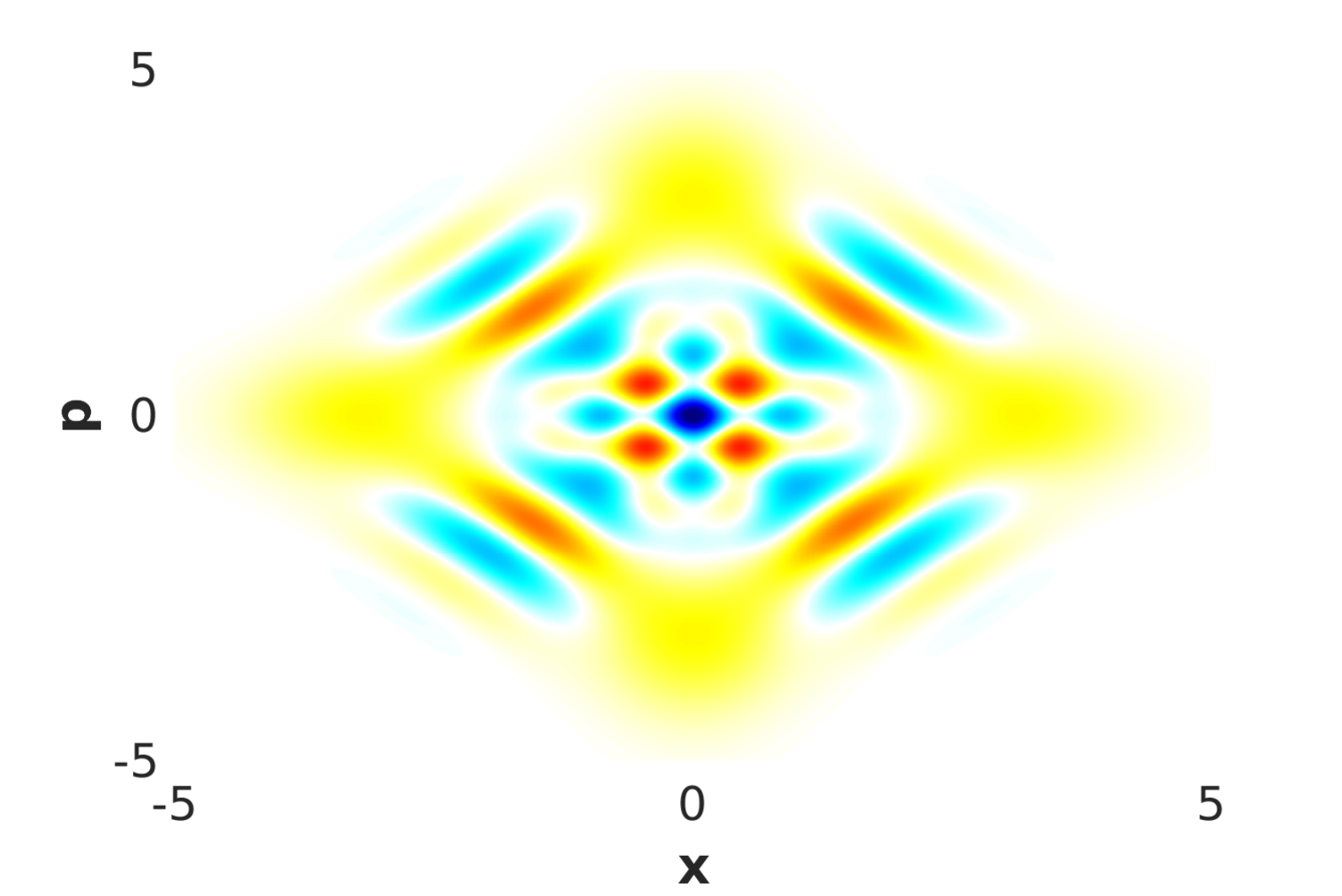}
\end{minipage}
\hspace{0.2cm}
\begin{minipage}[b]{0.45\linewidth}
\centering
\includegraphics[width=1.2\textwidth]{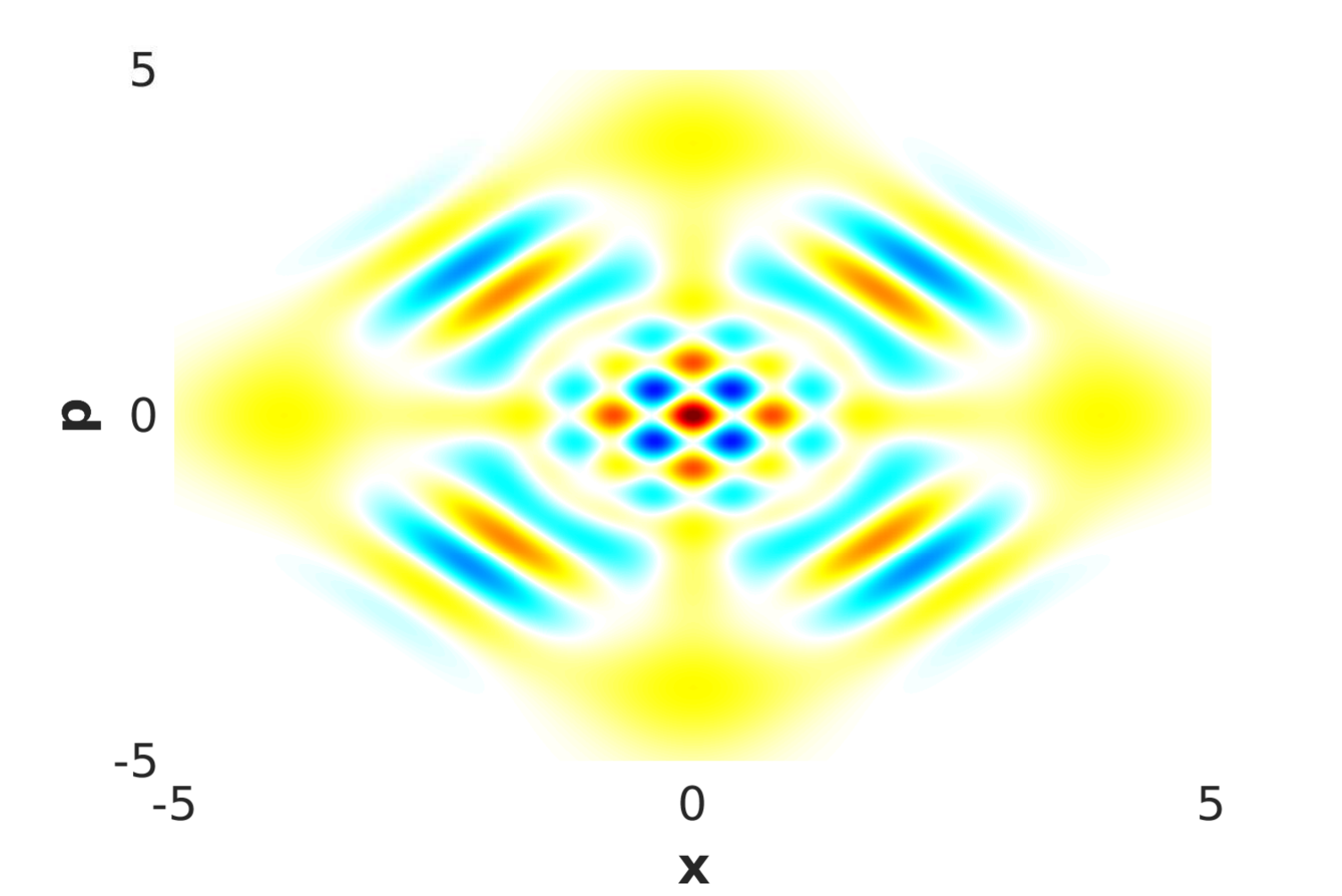}
\end{minipage}
\caption{(Color online) Appearance of sub-Planck structure in the phase space distribution of SSNS for squeezing parameter r=0.45 with Fock number n =1 (Top Left),\,  n=2 (Top Right), \, n=3 (Bottom Left), \, n=4 (Bottom Right).}
\label{2ndstate}
\end{figure}

Fig.\ref{firststate} and Fig.\ref{2ndstate} depict the phase space distribution of two states SSDNS and SSNS for different Fock number `$n$'. It is quite evident from the phase space distribution in Fig.\ref{firststate} and Fig.\ref{2ndstate} that both states present sub-Planck scales for arbitrary parameters ($r$ and $\alpha$) and oscillatory behaviour around the origin in the phase space. First state has clear and regular pattern as can be seen in Fig.\ref{firststate} for Fock number $n=1$ and $2$, while for higher number states, its structure does not show discernible oscillations. This can be explained as structure does not reveal in both direction until parameters squeezing and displacement are tuned to particular values optimized for the compass state's phase space behavior. Further, we explain the motivation for the proposal of the second state including only squeezing acting on the number state ($S[re^{i\phi}]\ket{n}$). Direction of squeezing of the number state can be controlled in the phase space with tuning of squeezing angle ($\phi$). Therefore, to obtain similar phase space distribution to the compass state, two operation of the equal amount squeezing with squeezing angle difference $\pi$,  are required in the phase space. This leads to the proposition with superposition of two squeezing operators acting on the Fock state `$n$'. This SSNS's  Wigner function ($W_{\psi_{2}}(x,p)$) clearly shows sub-Planck oscillation behavior in Fig.\ref{2ndstate}. Due to presence of the higher order 2D Hermite polynomial in the Wigner function of $\psi_{1}(x)$ and $\psi_{2}(x)$, it becomes difficult to extract the oscillation frequency of the phase space interference. We know that result of these oscillatory structure in the phase space leads to distinguishable state when state suffers small shifts in the phase space because of external perturbation. Comparing the first zeros of the overlap between unperturbed $\psi_{1}(x)\,(\psi_{2}(x))$ and perturbed state $\psi_{1}(x+ \delta)\,\left(\psi_{2}(x+\delta)\right)$, allows us to list the degree of sensitivity against the quadrature fluctuations ($\delta=\delta x+\iota \delta p$) for different parameter $r$ and $\alpha$. In the next section, we obtain the fidelity of both states with the compass state (KS) for few variation of the squeezing and displacement parameters. 

%%%%%%%%%%%%%%%%%%%%%%%%%

%%%%%%%%%%%%%%%%%%%%%%%%%

%%%%%%%%%%%%%%%%%%%%%%%%%

\section{Fidelity and Small shift sensitivity}\label{sec:IV} To quantify similarity in terms of distance,
we now compute the fidelity of states $\psi_{1}$  and $\psi_{2}$ with KS for large amplitude $(|\beta|>0.5)$ such that there exists small area phase-space structures along both field quadratures. Fidelity ($\mathcal{F}_{\psi}$) is, a measure for determining how two states are close, an overlap between the target state and acquired state. This measure $\mathcal{F}_{\psi}$ for the proposed state $\ket{\psi}$ with KS $\ket{\psi^{l}_{ks}(\pm)}$ in the terms of squeezing and displacement parameter is given as
$$\mathcal{F}_{\psi_{1}}^{(l,\pm)}=\left|\bra{\psi_{1}}\ket{\psi^{l}_{ks}(\pm)}\right|^{2}=N_{1}^{2}\left|O_{\pm}^{l}(r,\alpha)+O_{\pm}^{l}(r,-\alpha)\right|^{2}$$
and
$$\mathcal{F}_{\psi_{2}}^{(l,\pm)}=\left|\bra{\psi_{2}}\ket{\psi^{l}_{ks}(\pm)}\right|^{2}=N_{2}^{2}\left|O_{\pm}^{l}(r,0)+O_{\pm}^{l}(-r,0)\right|^{2},$$  

where symbols used are
\begin{multline*}
O_{\pm}^{l}(r,\alpha) = N_{3}^{2}\left(I_{n0}(r,0,\alpha,\beta)\pm (-1)^{-l}I_{n0}(r,0,\alpha,-\beta)+\right.\\\left. (-i)^{-l}I_{n0}(r,0,\alpha,i\beta)\pm (i)^{-l}I_{n0}(r,0,\alpha,-i\beta)\right) 
\end{multline*}
% and
% \begin{multline*}
%     f_{1}(r,\alpha,\beta)=\frac{e^{-\Im(\beta^{*} )^2}\sqrt{e^{r}}}{\sqrt{\pi } \sqrt{2^n n!}}\int_{-\infty}^{\infty}\,H_{n}(e^{r}x-\sqrt{2}\alpha)\\\times e^{-\frac{(x-\sqrt{2}\beta^{*})^{2}}{2}}{e^{-\frac{(e^{r}x-\sqrt{2}\alpha)^{2}}{2}}}dx
% \end{multline*}
with $N_{3}$\, being normalization of the KS.

\begin{table}[H]
 \caption{\label{tab:my_label} Fidelities of SSDNS and SSNS.}
   \centering
   \begin{ruledtabular}
    \begin{tabular}{cccccc}
    \textnormal{State $\mathcal{F}^{(l,\pm)}_{\psi}$} &$\alpha$  & $n$& $r$& KS $(\beta)$ & Fidelity ($\mathcal{F}^{(l,\pm)}_{\mathbf{\psi}}$) \\ \hline
    \vspace{0.1cm} \footnotesize{\textnormal{$\mathcal{F}^{(3,+)}_{\psi_{2}}$}}&0  & 3& 0.2& $1.41$& 0.9998  \\
    \vspace{0.1cm} \footnotesize{\textnormal{$\mathcal{F}^{(2,+)}_{\psi_{2}}$}} &0  & 2& 0.3& $1.41$& 0.9997  \\ 
    \vspace{0.1cm} \footnotesize{\textnormal{$\mathcal{F}^{(1,+)}_{\psi_{2}}$}} &0  & 1& 0.3& $1.01$ & 0.9994  \\
    \vspace{0.1cm} \footnotesize{\textnormal{$\mathcal{F}^{(1,+)}_{\psi_{2}}$}} &0  &0& 0.4& $0.81$ & 0.9998  \\
    \vspace{0.1cm} \footnotesize{\textnormal{$\mathcal{F}^{(0,-)}_{\psi_{1}}$}} &$0.61$& 1  &0.15 &$0.7(1+i)$&0.9995  \\
    \vspace{0.1cm} \footnotesize{\textnormal{$\mathcal{F}^{(0,-)}_{\psi_{1}}$}}&$0.7$ &1  &0.2 &$0.85(1+i)$&0.9960\\ 
    \vspace{0.1cm} \footnotesize{\textnormal{$\mathcal{F}^{(-1,-)}_{\psi_{1}}$}}&$0.5$& 2  &0.49 &$0.9(1+i)$&0.9634
    \end{tabular}
    \end{ruledtabular}
   \label{fid-tab}
\end{table}

The precision measurement processes are dependent on the energy resources, for example average of photon  number  $\langle n\rangle$ of the probe state. In standard quantum limit (SQL), small displacement measurements are independent of the average of photon number, using coherent state as a probe, while presence of sub-Planck structures in the probe state (such as cat state and compass state of coherent amplitude $\beta$) allow measurement of small displacement ($\delta=\sqrt{2}(\delta x +i \delta p)$) inversely proportional to coherent amplitude $|\beta|$ (where $|\beta| \sim \sqrt{\langle n\rangle}$ with $|\beta|>2$ for the cat and KS state), leads to Heisenberg limited sensitivity (HL).  We know from the Table.\ref{fid-tab} of fidelities for both proposed states with KS that tuning to particular values of squeezing $`r'$ and displacement $`\alpha'$ parameters for Fock number $n=1$ provide overlap beyond 99$\%$. For SSDNS with $n=1$, we find great fidelity for different coherent amplitude $\beta$ of the KS $\left(\ket{\psi_{ks}^{l}(-)}\right)$, while other values of $n$ for this state does not make good candidate for the comparison with KS on the basis of overlap ($\mathcal{F}_{\psi}$). Our second state (SSNS) for every Fock number ($n=0,1$ and so on), shows fidelity beyond 99$\%$ with the KS $\left(\ket{\psi_{ks}^{l}(+)}\right)$ for different coherent amplitude ($\beta$).

Further, use of these states in the measurement of small shifts in phase space due to perturbation and weak forces external to the system, we study overlap of these states with those having small amount of displacements in phase space. Overlap ($O_{\delta}$) of perturbed and unperturbed state of system describes distinguishablity of state distribution against small shifts in phase space and quantifies sensitivity to the shifts $\delta$ as $O_{\delta}\longrightarrow0$ with $\delta\longrightarrow \delta_{0}$ (first zero of $O_{\delta}$). Smaller the overlap $$O_{\delta}=2\pi\int_{-\infty}^{\infty}dxdp\, W_{\psi}(x,p)W_{\psi}(x+\delta x, p+\delta p)$$$$=\left|\bra{\psi}D[\delta]\ket{\psi}\right|^{2}$$ for the state $\ket{\psi}$, higher the sensitivity against small perturbations ($\delta x,\,\delta p$) in the
phase space. This overlap of system state, can be found through measurement of the two level system (TLS), interacting with the probe state ($\ket{\psi}$) of the system, either in upper state ($\ket{e}$) or lower state ($\ket{g}$).

In measurement strategy of small displacements, a probe state of the oscillator (field state or ion state) entangled with TLS is prepared through appropriate time unitary evolution $U$ by initializing system in the product state of oscillator ($\ket{\phi_{i}}$) and TLS ($\ket{e}$). After the preparation, small displacement operator $D[\delta]$ as a consequence of weak forces or small perturbations external to the composite system (oscillator and TLS) is applied along with reversing the action of $U$ thereafter. Final state obtained of this composite system \cite{Hl-zurek}, is written as 
$$\ket{\psi_{f}}=U^{-1}D[\delta]U\ket{\phi_{i}}\ket{e}=\sqrt{p_{e}}\ket{\Phi_{1}}\ket{e}+\sqrt{p_{g}}\ket{\Phi_{2}}\ket{g}.$$

The $U$ unitary operator is such that probe state $\ket{\psi}=U\ket{\phi_{i}}$ and the excited state probability $$p_{e}=\left|\bra{\phi_{i}}\ket{\psi_{f}}/\bra{\phi_{i}}\ket{\Phi_{1}}\right|^{2}\sim \left|\bra{\psi}D[\delta]\ket{\psi}\right|^{2}=O_{\delta}.$$ We can take the advantage of preparing probe state $\ket{\psi}$ into either one of the cat state, compass state (KS), SSDNS and SSNS. In the measurement of two level atom in the excited state, obtained probability function ($p_{e}=p_{e}(\delta)$) is inverted to find small shift $\delta$. To achieve estimated small displacement very close to its true value, measurement of TLS of the composite system, in its excited or ground state is repeated $R$ times. Consider $m$ times ($m<<R$) atom is found to be in excited state. Probability of this measurement is written \cite{luis, Hl-zurek} as $$\frac{R!}{m!((R-m)!)}p_{e}^{m}(1-p_{e})^{R-m},$$
where this binomial can be approximated to normal distribution, when  number $R$ is very large as compared to $m$. With prior knowledge of $0<\delta<\delta_{0}<<1$ and true value $s=\delta^{2}$, normal distribution in terms of $R,\,p_{e}$ is
$$ \frac{1}{\sqrt{2\pi\Delta^{2}s }}e^{-\frac{(s-\bar{s})^{2}}{2\Delta^{2}s}},$$
 where $\bar{s}$ and $\Delta^{2}s$ are the estimated value of shifts and variance related to this distribution.
 
\begin{center}
    \begin{figure}[H]
        \centering
        \begin{tabular}{ccc}
    \includegraphics[scale=0.45]{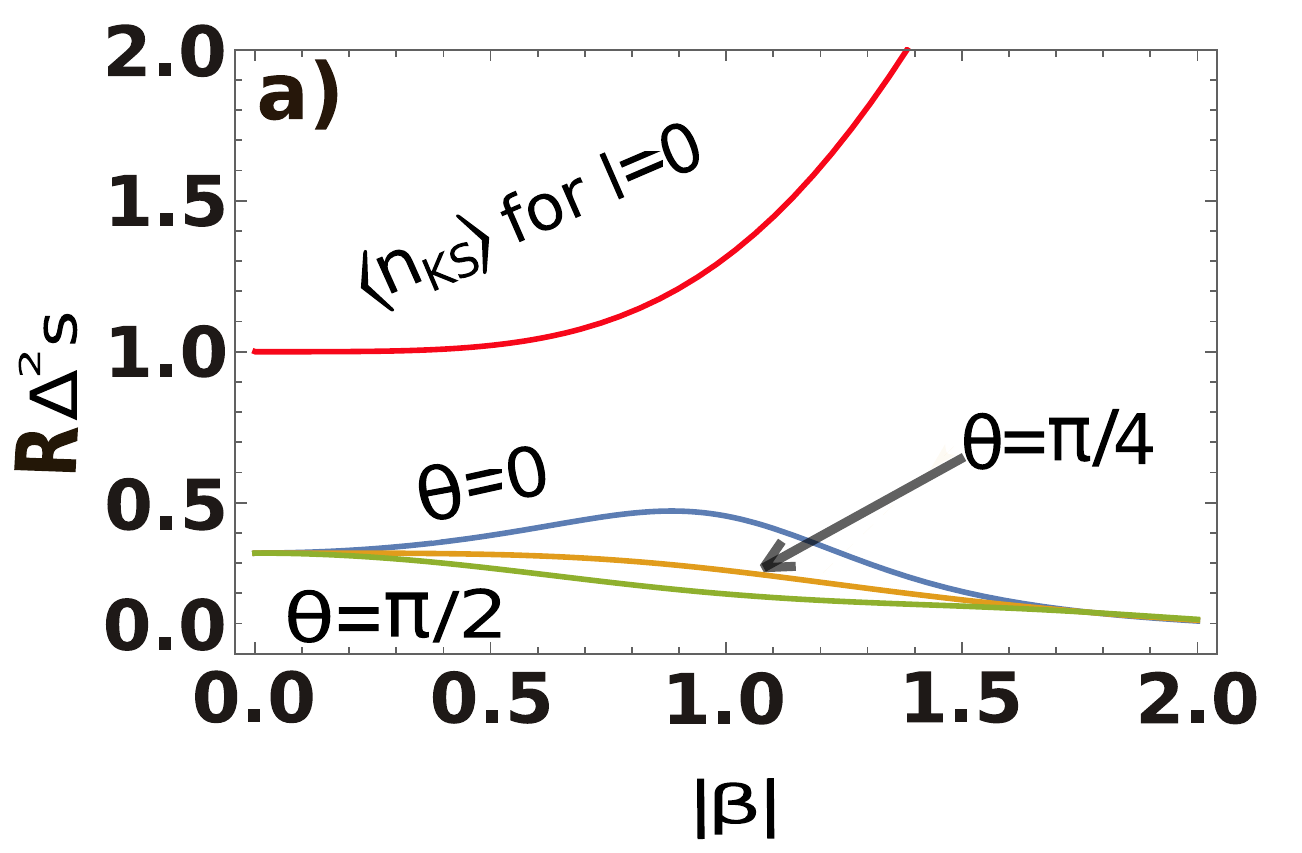} & \includegraphics[scale=0.45]{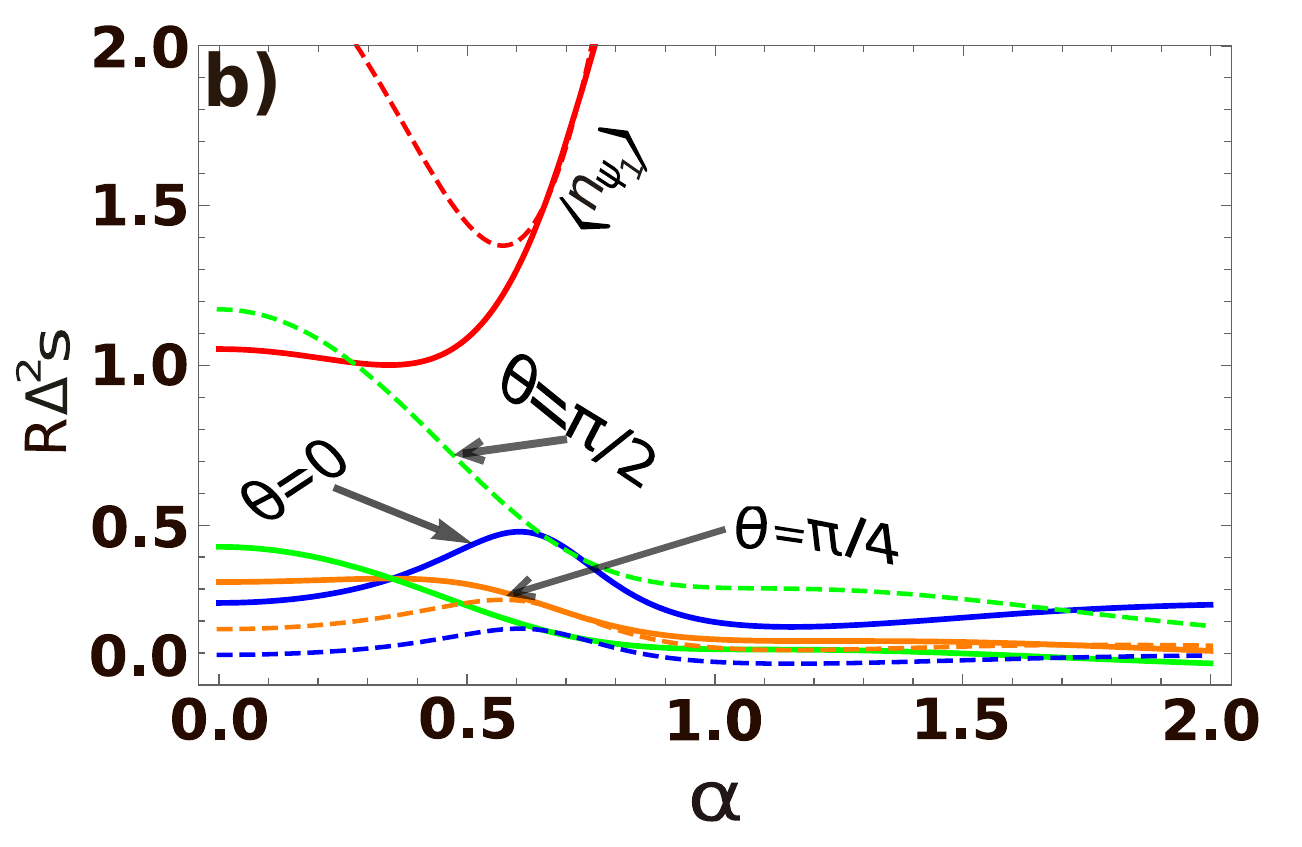} &      \includegraphics[scale=0.45]{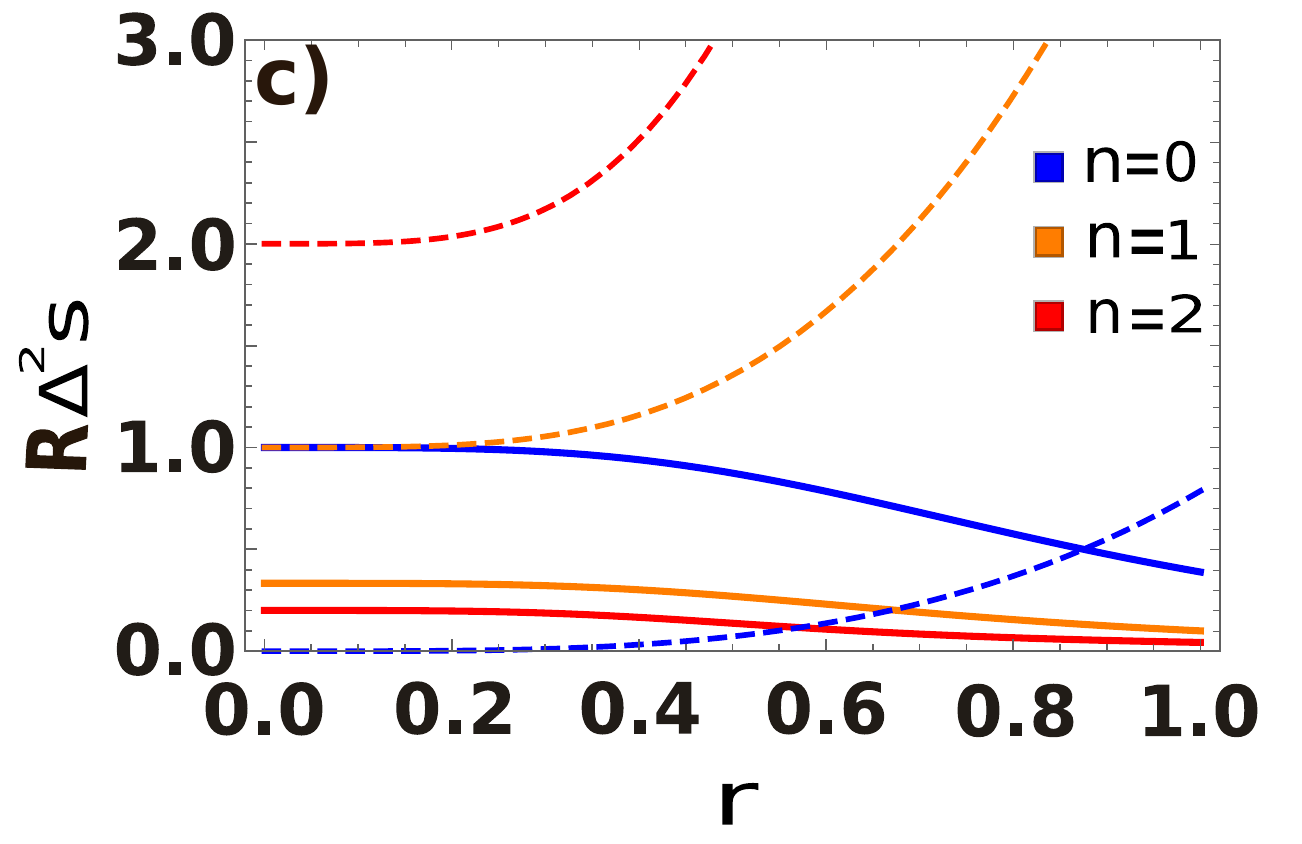} 
    \end{tabular}
         \caption{(Color online) 
         Variances $R\Delta^{2}s$ and average photon numbers $\langle \hat{n}\rangle$ are plotted for states (a) $\ket{\psi_{KS}^{0}(-)}$ (b) $\ket{\psi_{1}}$  (c) $\ket{\psi_{2}}$ describing small phase-space displacement $(s)$ detection. Plots (a) and (b) include red curves describing average photon numbers, while others are for variances. Variances in plots $(a)$ and $(b)$ depend upon direction ($\theta$) of the displacements, showing behavior of conjugate quadrature akin to the position-momentum inequality. Dashed and thick curves in plot $(b)$, are illustrated for the squeezing $r=0.13$ and $0.63$ with the Fock number $n=1$,  respectively. In the third plot $(c)$, respective dashed and thick curves depict  $\langle\hat{n}\rangle$ and $R\Delta^{2}s$ for the different Fock numbers $n=0,1,2.$        
         Finally, all Plots $(a-c)$ clearly describe the inverse correlation between  $R\Delta^{2}s$  and $\langle \hat{n}\rangle$ for small displacement $(s)$ detection along different directions $\theta =0,\, \pi/4,\,\pi/2$ in quadrature phase-space, with $R$ repetitive measurements.}
         \label{kssensitvity}
    \end{figure}
 \end{center}

\begin{center}
    \begin{figure}[H]
     \begin{tabular}{ccc}
    \includegraphics[scale=0.45]{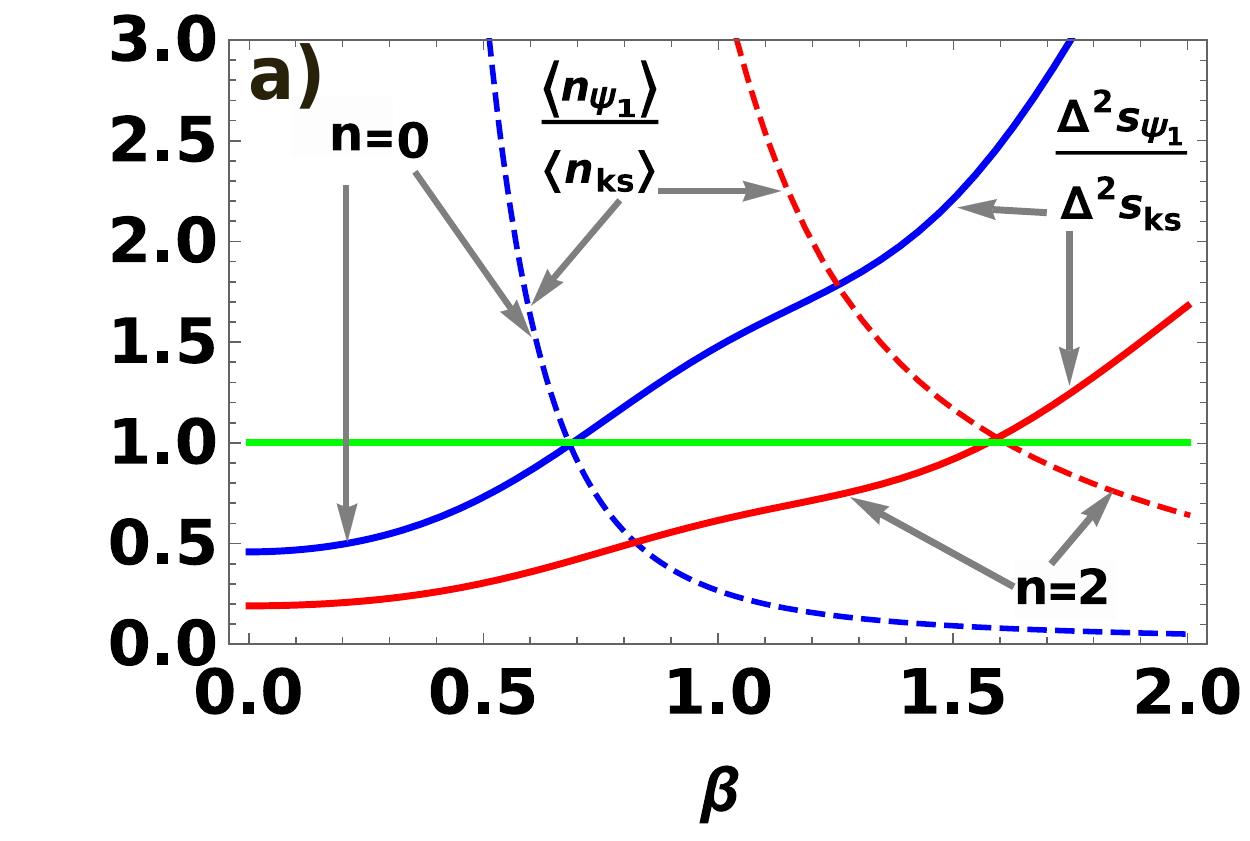} & \includegraphics[scale=0.45]{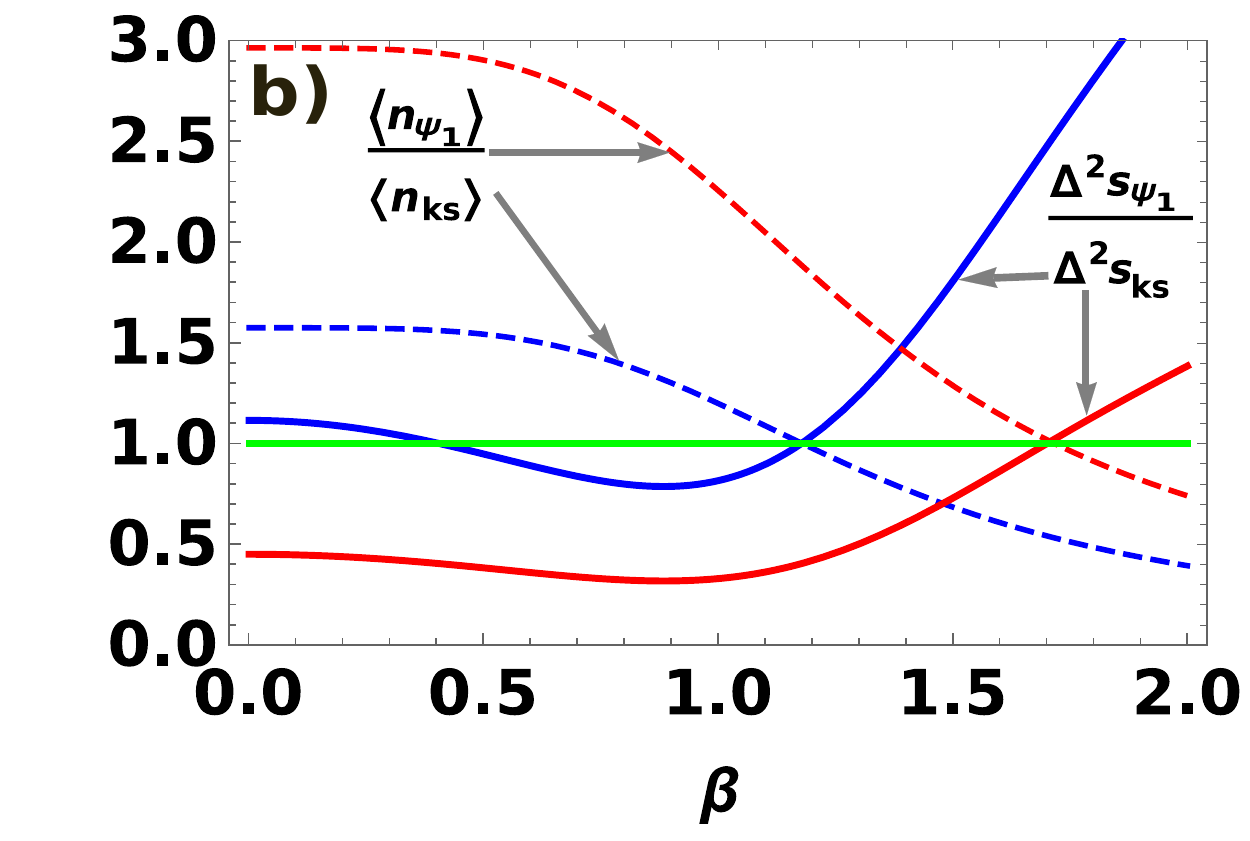} &      \includegraphics[scale=0.521]{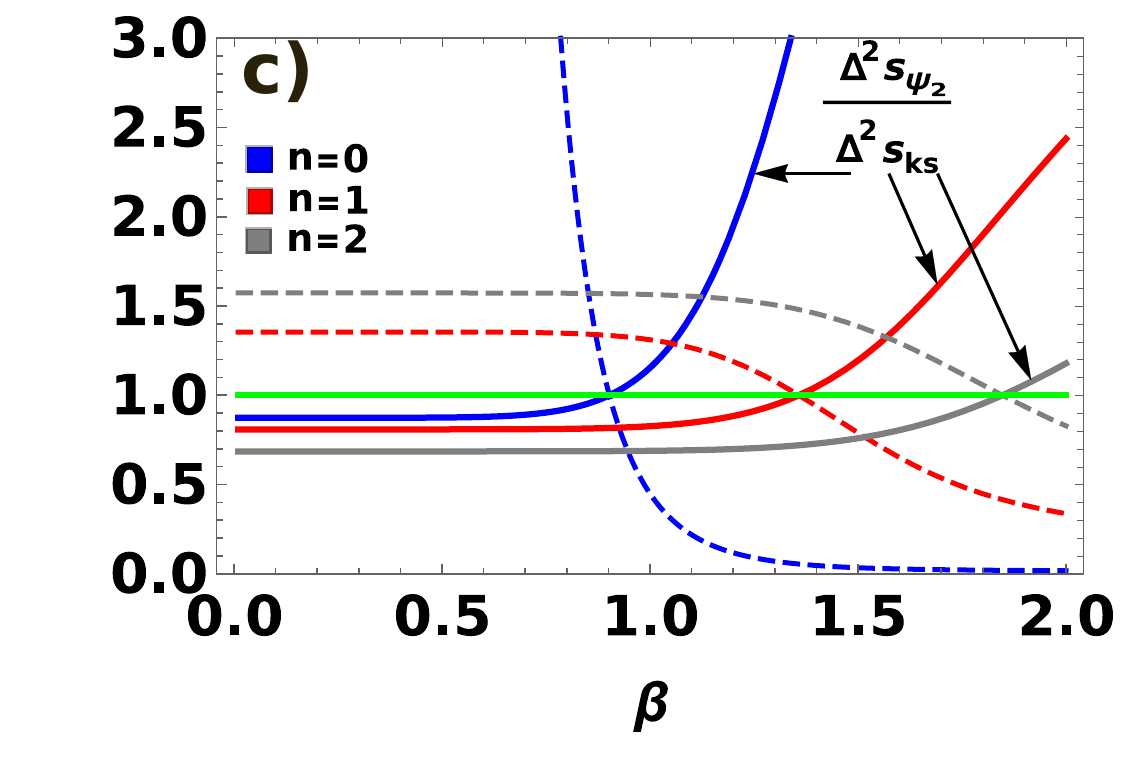} 
    \end{tabular}
    % \begin{minipage}[b]{0.3\linewidth}
    % \centering
    % \includegraphics[width=1.07\textwidth,height=.9\textwidth]{Main-tex/Images/ration02vari1ststate.pdf}
    % \end{minipage}
    % \hspace{0.05cm}
    % \begin{minipage}[b]{0.3\linewidth}
    % \centering
    %  \includegraphics[width=1.07\textwidth,height=.9\textwidth]{Main-tex/Images/ration1vari1ststatedpi.pdf}
    % \end{minipage}
    % \hspace{0.05cm}
    % \begin{minipage}[b]{0.3\linewidth}
    % \centering
    %  \includegraphics[width=1.07\textwidth,height=.9\textwidth]{Main-tex/Images/ration012vari2ndstate.pdf}
    % \end{minipage}
    \caption{(Color online) Plots (left, middle, right) depict ratio of average photon numbers $\langle n\rangle$ (dashed) and variances $\Delta^{2}s$ (thick) for $\ket{\psi_{1}}$ or $\ket{\psi_{2}}$ state with that of KS state ($\ket{\psi^{l}_{ks}(-)}$ or $\ket{\psi^{l}_{ks}(+)}$) against coherent magnitude $|\beta|$ of KS state, respectively. Left (a) shows both ratios for $\psi_{1}$ state with even Fock number (n=0, 2) and $\psi^{1}_{ks}(-)$ state. Blue (Red) curves are plotted for parameters $r= 0.796\, (0.214)$,$\,n=0\,(2)$ and $\alpha=0.815\, (0.68)$. Middle (b) has ratios illustrating blue (red) curves plotted for parameters  $r= 0.2\,(0.3)$ and $\alpha=0.7\,(1.5)$, with $n=1\,(1)$ of $\psi_{1}$ state and  $\psi^{0}_{ks}(-)$.   With adjusting values of parameter $`r'$ and $`\alpha'$, dashed and thick curves show region $>$ and $<1$, describing inverse proportionality behavior of average photon number $\langle a^{+}a\rangle$ and variance. Similar behavior is seen in the rightmost (c) plot depicting ratios for different Fock number $n=0,1,2$ of $\psi_{2}$ state and parameters $l=0,1,2$ of $\psi_{ks}^{l}(+)$ with squeezing parameter $r=0.5$.}
\label{numdist1}
\end{figure}
\end{center}

 Smaller the variance, closer the estimator ($\bar{s}$) to true value ($s$) as seen in the above normal distribution. We calculate variances $\Delta^{2}s$ and average number $\langle a^{+}a\rangle$ for $\psi_{1}$, $\psi_{2}$ and $\psi_{KS}$ as probe states, against variation in squeezing ($r$), displacement ($\alpha$) and Fock number ($n$) along with the amplitude $|\beta|$ of the KS. We approximated the variance to $\delta^{2}/(R\,c)<\delta_{o}^{2}/(R\,c)<1/(R\,c),$\, with $c$ constant dependent on parameters of the probe state. We plot variance $\Delta^{2}s = 1/\left(R\, c\right)$ and average photon number $\langle a^{+}a\rangle$ for the compass state $\left(\psi_{KS}\right)$, SSDNS $\left(\psi_{1}\right)$ and SSNS ($\psi_{2}$), in the Figs.\ref{kssensitvity}-\ref{numdist1} to study their cost in perturbation sensitivity. 
 
 Plots in Fig.\ref{kssensitvity} are shown for the variance and average number against parameters such as coherent amplitude $(|\beta|)$, displacements $(\alpha)$ and squeezing $(r)$ for states $(a)\, \ket{\psi_{KS}^{0}(-)},\,(b)\, \ket{\psi_{1}}\,$ with $n=1$ and $(c)\, \ket{\psi_{2}}$, respectively. Inverse relationship between variance and average number is clearly evident from all these plots in Fig.\ref{kssensitvity}.  

 It is known that variance is inversely proportional to average photon number $\langle n\rangle$ for KS state \cite{Hl-zurek} in the weak perturbation detection. This relation between variance and $\langle n\rangle$ is clearly seen for SSDNS and SSNS in Fig.\ref{kssensitvity}. To further analyse that SSDNS and SSNS ($\psi$) have similar behavior of $\Delta^{2}s$ and $\langle n\rangle$ with KS, we plot the ratio of variances ($\Delta^{2}s_{\psi}/\Delta^{2}s_{\psi_{KS}}$) and average number $\langle n\rangle_{\psi}/\langle n\rangle_{\psi_{KS}}$ in the Fig.\ref{numdist1}. In the Fig.\ref{numdist1}, plot on the left shows the variance ratio $\Delta^{2}s_{\psi_{1}}/\Delta^{2} s_{\psi_{KS}^{1}(-)}$ (thick) and average number ratio $\langle n_{\psi_{1}}\rangle/\langle n_{KS}\rangle$ (dashed) for SSDNS $\psi_{1}$ with $n=0\,,2$ and KS $\psi_{KS}^{1}(-)$. For appropriate parameters $r$ and $\alpha$, variance and average number cross each other for ratios equal to $1$ (green line) signifying same precision measurement of small shift $\delta$ at equal energy cost of the both probe states. Similar behavior is seen for SSDNS with $n=1$ and KS $\psi_{ks}^{0}(-)$ in the Fig.\hyperref[numdist1]{5b}. %Increase in the $\beta$ at the $r=0.5$ on the rightmost
 Also the last Fig.\hyperref[numdist1]{5c} for the $r=0.5$, shows the crossing of the variances (thick) and average numbers (dashed) ratios for the Fock numbers $n={0,1,2}$ of SSNS corresponding to respective values $l={0,1,2}$ of KS $\psi_{ks}^{l}(+)$. All these plots in Fig.\ref{numdist1} describe small phase space fluctuation measurements correspond to same HL sensitivity with appropriate choice of parameters for both proposed states when compared to the KS.
 
 % Figs describe plot for the overlap of the first state (SDNS) against small phase space shift ($\delta s$). Curves having fidelity over 99$\%$ shows symmetric shift along x and p when overlap is zero, while others for small fidelity have asymmetric shift along x and p as consequence of less or more squeezing present in first state. Increase in asymmetry of the shift ($\delta s$) and decrease in amount of the $\delta s$  can be seen with increase in $`n'$ the Fock number excitation in the first state. 

 \onecolumngrid
 
\twocolumngrid
%%%%%%%%%%%%%%%%%%%%%%%%%%%%5
\section{Use of probe in the damping constant estimation}\label{sec:V}
\begin{center}
\begin{figure}
    \centering
    \includegraphics[width=6.0 cm,height=4.0 cm]{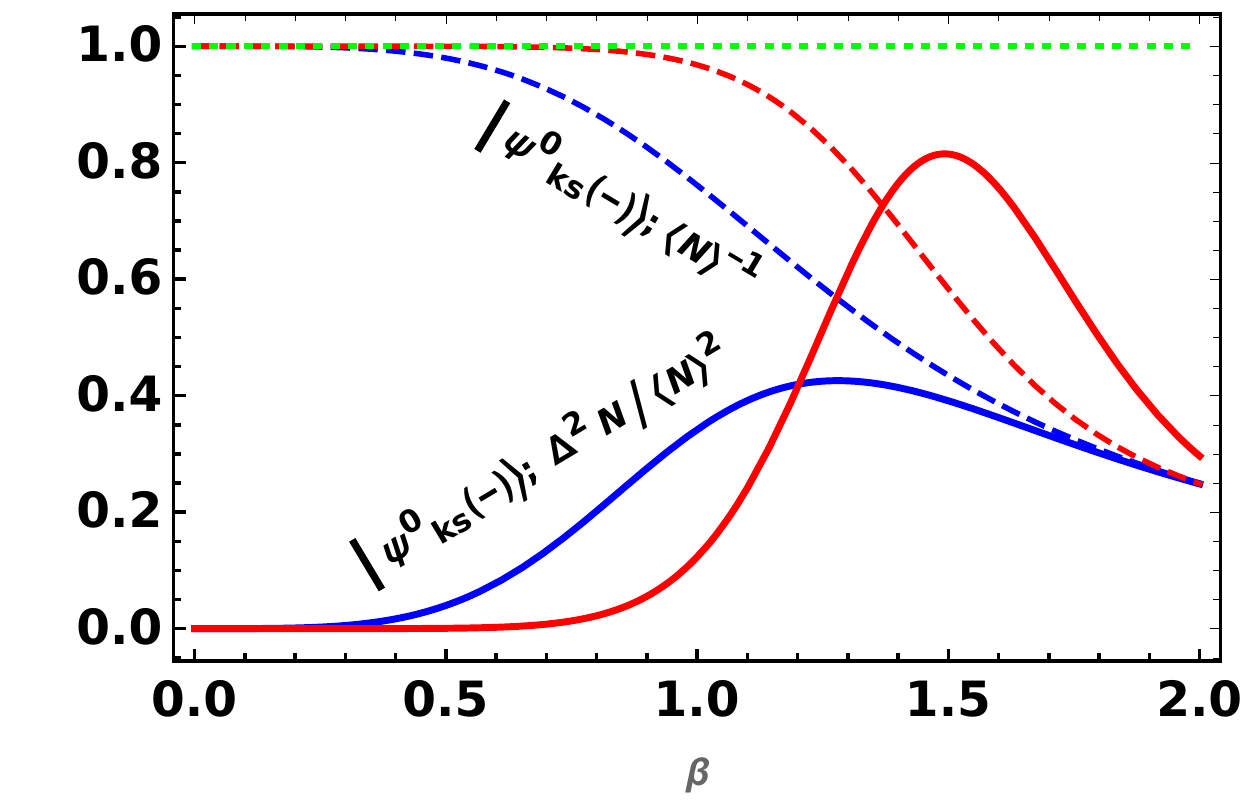}
     \caption{(Color online) Ratio of number variance and squared average number $\frac{\Delta^{2} N}{\langle N\rangle^{2}}$ (Thick lines) is plotted with comparison to the inverse of average photon number $(\langle N\rangle^{-1})$ (Dashed lines) for states $\ket{\psi^{0}_{ks}(-)}$ (Blue) and  $\ket{\psi^{1}_{ks}(+)}$ (Red), respectively. Parameter $\beta$ is the coherent amplitude of the KS. Both plotted quantities for ks are near to that of single photon state, when $\beta$ is small in magnitude.    }    \label{dampest}
\end{figure}
\end{center}

In the open system, light interacting with the environment such as absorption media, passing through beam splitter (BS) causes its damping and leads to the decoherence of quantum state. In this process of coherence loss, it allows to gain the information about damping parameter of the medium. Decay of the light amplitude in the cavity is described by the master equation:
$$\frac{\partial \rho}{\partial t}= \kappa (2 a\rho a^{+}- \{a^{+}a,\rho\}),$$
where $\kappa$ is the damping constant of the cavity. This equation also describes dynamics of BS, opening the possibility of estimating either reflectivity ($1-\eta$) or transmitivity ($\eta=e^{-2\kappa t}$) of BS. Use of the quantum state of light  being sensitive to the small changes in the parameter ($\kappa$), can increase precison of estimating parameter $\kappa$. Error ($\Delta \kappa$) associated in measurement of $\eta$, obtained through output of BS \cite{agarwal-damp}, can be found from error-propagation method:
$$ \Delta\kappa=\frac{\sqrt{\Delta^{2}\hat{N}_{out}}}{\left|\partial \langle\hat{N}_{out}\rangle/\partial\kappa\right|}= \sqrt{\frac{\Delta^{2}N_{in}}{4t^{2}\langle \hat{N}_{in}\rangle^{2}}+\frac{(\eta^{-1}-1)}{4t^{2}\langle \hat{N}_{in}\rangle}}, $$
where $\langle \hat{N}_{out}\rangle$ ($\langle \hat{N}_{in}\rangle$) and $ \Delta^{2}\hat{N}_{out}$ ($ \Delta^{2}\hat{N}_{in}$) are average number and number variance, respectively, at the output (input) for the probe passing through BS for $t$ time. As is known, for optical probe state, Fock states $|n\rangle$ provide lowest error bound as the number variance is zero with cost of n average photon number. Hence, Use of single photon state leads to lowest bound with least cost of 1 ($\langle a^{+}a\rangle$). We know from Fig.\ref{numdist}, with increase in the superposition number of different coherent state (for example $\sum_{k=0}^{max}\ket{\alpha_{k}}$), there is corresponding larger gap for adjacent non-zero probability in the Fock space. It leads to decrease in number fluctuation and therefore close to the 
number state $\ket{n}$. $\Delta\kappa$ depending on the variance and average reaches to minimum for small magnitude $\beta$ of the $\ket{\psi^{0}_{ks}(-)}$ and $\ket{\psi^{1}_{ks}(+)}$ states similar to single photon state clearly evident in Fig.\ref{dampest}.  Therefore, these two states as the probes show great potential in estimating damping constant of the BS. 

\section{Preparation of the state} We now focus on the generation of the proposed states in the optical setup. One procedure described in detail \cite{stateprep-vogel} for preparation of any single mode radiation field's state can be used in our case of both proposed states. This method requires $N$ two level atoms and high $Q$-cavity such that damping of photons and spontaneous emission rate of atoms in the cavity are negligible. The cavity is initialized in vacuum state while allowing atoms prepared in superposition states of two levels beforehand, to pass through the cavity one by one. Atoms interact via JC-Hamiltonian with cavity field state, increasing Fock state superposition by one, when found in the ground state as they leave the cavity. By tuning parameter d in two level atom prepared state $\,\ket{e}+d\ket{g}\,$ allows to effectively control the amplitude of the Fock state superposition formed in the cavity and hence leading to the desired state preparation after $N$ atoms enter and leave the cavity one by one. To obtain maximum overlap of the formed state in cavity with desired state, $N$ must be greater than average of the photon number for the desired state. It can also be decided from the probability number distribution as trend reaches appreciably small after obtaining the desired state. 

Light-matter interactions have allowed generation of non-classical states of light within RWA and beyond RWA. Extensive amount of theoretical research has been performed in solving dynamics of JC-Rabi model, but realization of this model in and beyond ultra strong coupling (USC) has shown itself quite challenging to reach in cavity QED. Advances in technology, has allowed uses of superconducting qubit in circuit QED to simulate dynamics of Rabi model in USC regime, as proposed \cite{pro-cir0} and reported recently \cite{ex-cir0,ex-cir1,ex-cir2,ex-cir3}. Hamiltonian for superconducting qubit-resonator interaction describing Rabi model in circuit QED in the presence of two-photon driving is:

\begin{multline*}
    \mathcal{H}_{1}=\omega_{o} a^{+}a + \frac{\omega_{a}}{2} \sigma_{z} +g(a^{+}+a)\sigma_{x}\\ +G({a^{+}}^{2}e^{-2i\omega_{t} t}+a^{2}e^{2i\omega_{t} t}), \end{multline*}

where $\omega_{o}, \omega_{a}$ and $\omega_{t}$ are resonator mode, qubit frequency and two-photon drive's frequency. Parameters $g$ and $G$ describe interaction strength and two-photon drive strength. This Hamiltonian $\mathcal{H}$ further can be modified by using squeeze $S[\chi]=exp(\frac{-\chi{a^{+2}}+\chi^{*}a^{2}}{2})$ and displacement $D[\lambda]=\exp(\lambda a^{+}-\lambda a)$ unitary operators \cite{rab-agrw}. Diagonalized Hamiltonian reads as
\begin{multline*}
\mathcal{H}_{d}=e^{i\omega_{t} t a^{+}a}D^{+}[\lambda \sigma_{x}]e^{-i\omega_{t} t a^{+}a}S^{+}[\chi]e^{i\omega_{t} t a^{+}a}\left(H\right.\\\left.-\omega_a \sigma_{z}/2\right)e^{-i\omega_{t} t a^{+}a}S[\chi]e^{i\omega_{t} t a^{+}a}D[\lambda \sigma_{x}]e^{-i\omega_{t} t a^{+}a}\\= \sqrt{\Lambda^{2}-4G^{2}} a^{+}a +\frac{1}{2}(\sqrt{\Lambda^{2}-4G^{2}}-1)\\-\frac{2g^{2}}{\sqrt{\Lambda^{2}-4G^{2}}}\left(\frac{\Lambda -2G}{\Lambda +2G}\right)^{2}, \end{multline*}
where parameters detuning ($\Lambda$), phase space displacement ($\lambda$) and squeezing coefficient ($\chi$) for diagonalization of $\mathcal{H}_{1}$ (when $\omega_a =0$) are
\begin{multline*}
    \Lambda=\omega_{o}-\omega_{t},\, \lambda=\frac{g(\Lambda -2G)}{(\sqrt{\Lambda^{2}-4G^{2}})(\Lambda +2G)},\\ \chi=-\frac{1}{2}\log\left(\frac{\Lambda +2G}{\Lambda -2G}\right). \end{multline*}
 In the presence of degenerate qubit (i.e. $\omega_{a}=0$), above Hamiltonian $(\mathcal{H})$ can be diagonalized in the squeezed displaced number state $'S(\chi)[D(\lambda)\pm D(-\lambda)]\ket{n,\pm}'$ in the even (odd) parity branches, as discussed in \cite{eig-rabi2nd}. Introduction of qubit with $\omega_{a}\neq 0$, lifts degeneracy and its eigen-spectrum can be approached through first order perturbation \cite{eig-rabi1st}, considering qubit $\sigma_{z}$ term as perturbation, where this regime described as perturbative-deep strong coupling. In this regime for $\omega_{a}\leq \sqrt{\Lambda^{2}-4G^{2}}$ \cite{eig-rabi2nd,eig-rabi1st}, above found eigenstates have 99$\%$ overlap with exact numerically simulated states in \cite{eig-rabi2nd} with lifted degeneracy. Generation and properties of the first proposed state, has also been theoretically discussed on the ion trap platform \cite{fasal}.

 Production of SSNS can be availed with help of system containing two excitation emission and loss of resonator (oscillator) mode. Two photon rabi model has been theoretical interest and proposed in circuit QED, for example zero current bias in SQUID loop controls linear interacting term leading to two photon-qubit dominant term for SQUID-flux qubit \cite{tpra-mod} interaction. Use of Hamiltonian in the form, 
 $$\mathcal{H}_{2}=\left(g_{1}a^{+}a+g_{2}(a^{2}+{a^{+2}})\right)\sigma_{z} +\omega a^{+}a, $$
containing interaction term of two-quanta resonator mode  with superconducting qubit can be diagonalised in interaction picture. Time evolution of state ($\ket{\psi_{I}}=e^{ig_{1}t\sigma_{z}a^{+}a}\ket{\psi_{\mathcal{H}_{2}}}=U\ket{\psi_{\mathcal{H}_{2}}}$) is given by 

    \begin{multline*}\ket{\psi_{I}(t)}=e^{-i\int_{0}^{t}dt U\left(g_{2}(a^{2}+{a^{+}}^{2})\sigma_{z}+\omega a^{+}a\right)U^{-1}}\ket{\psi_{I}(0)}
    \\=e^{i g_{1}\sigma_{z} t a^{+}a/2}S(-r\sigma_{z})e^{-i\frac{t}{2}\left( (2a^{+}a +1) \sqrt{\omega^{2}-4{g_{2}}^{2}{j^{2}}_{o}(g_{1}t)} \right)}\\e^{i\frac{\omega t}{2}}S(r\sigma_{z})e^{-i g_{1}\sigma_{z} t a^{+}a/2}\ket{\psi_{I}(0)},
\end{multline*}
where the zeroth spherical Bessel function $j_{0}(t)=t^{-1}\sin{t}$ and squeezing parameter $r =\frac{1}{2}\log{\left(\frac{\omega -2g_{2}j_{o}(g_{1}t)}{\omega +2g_{2}j_{o}(g_{1}t)}\right)},$ with $g_{2}$ being constrained by the condition $2g_{2}\leq\omega$ for real $`r$'. Initializing resonator mode and qubit state $\ket{\psi_{I}(0)}=\ket{n,+}$, $\mathcal{H}_{2}$ system's state evolved to state at t time is
\begin{multline*}
    \ket{\psi_{\mathcal{H}_{2}}(t)}=\mathcal{N}S\left(-re^{-ig_{1}\sigma_{z}t}\sigma_{z}\right)\\S\left(re^{-i\hat{\mathcal{A}}t}\sigma_{z}\right)e^{-ig_{1}\sigma_{z} t a^{+}a}\ket{n,+}\\=\mathcal{N}\left[S\left(-re^{-ig_{1}t}\right)S\left(re^{-i\lambda_{+}t}\right)e^{-ig_{1} t n}\ket{n,e}\right.\\\left.+S\left(re^{ig_{1}t}\right)S\left(-re^{-i\lambda_{-}t}\right)e^{ig_{1} t n}\ket{n,g}\right],\end{multline*}
where $\mathcal{N}$ corresponds to normalization constant of the state and $\hat{\mathcal{A}}=g_{1}\sigma_{z}+2\sqrt{\omega^{2}-4g_{2}^{2}j^{2}_{o}(g_{1}t)}$ with eigenvalues $\lambda_{\pm}=\pm g_{1}+2\sqrt{\omega^{2}-4g_{2}^{2}j^{2}_{o}(g_{1}t)}$ satisfies $\hat{\mathcal{A}}\ket{e}\,(\hat{\mathcal{A}}\ket{g})=\lambda_{+}\ket{e}\,\left(\lambda_{-}\ket{g}\right)$. Measurement of qubit of the above evolved state in a $\ket{\pm}$ basis will lead to the required superposition of the squeezed number state. 
\section{conclusion}\label{sec:VII}
We compared the number distribution, Wigner function, fidelity and sensitivity of SSNS and SSDNS with compass state. Investigation of these properties has led to the finding that appropriate tuning of squeezing $`r$' and displacement $`\alpha$' shows close number distribution, similar Wigner function and maximum fidelity with respect to the assumed form of KS $\ket{\psi_{ks}^{l}(\pm)}$ states. Number distribution has zero and non-zero probability amplitudes for the KS depending on $l$ values. This leads to the different choices of KS to obtain larger overlap (Fidelity $F_{\psi}^{l,\pm}$) with SSNS and SSDNS. Phase space distribution of SSDNS ($\psi_{1}$) does not show large similarity for $n>1$ with KS, also evident from fidelity Table.\ref{tab:my_label} where $\beta$ parameter is not large. For SSNS ($\psi_{2}$) depending on squeezing and $n$ Fock number, clear interferometric phase space distribution is seen with large fidelity. There is an increase in $\beta$ of KS, corresponding to addition in $n$ at a step one.  Further, uses of KS in detecting small perturbation are no more better than SSDNS and SSNS when variance compared with average energy cost, evident from Fig.\ref{kssensitvity}-\ref{numdist1}. In the view of generation of single photon states, SSNS, SSDNS and KS provide same precision for the estimation of damping parameter of BS (see Fig.\ref{dampest}). We know that given the challenges present in obtaining large coherent amplitudes ($\beta$) of the KS, there are not many models to produce KS states directly either in optical platform except Kerr interaction  and using approximate methods such as conditional measurement in BS \cite{Thekkadath2020engineering,hastrup} or in the microwave regime. Therefore, considering their many implications, theoretical models are provided based on their accessibility in the optical as well as circuit-QED.
Sub-Planck structures in SSDNS and SSNS, and their co-relation to the compass state will enable different methods for the preparation of such states, which may find applications in the quantum metrology, sensing, communication and error correction. 

\begin{acknowledgements}
Arman is thankful to the University Grants Commission and Council of Scientific and Industrial Research, New Delhi, Government of India for Junior Research Fellowship at IISER Kolkata.
\end{acknowledgements}

\appendix*
\section{}\label{appendix:a}
We find $x$-projection of squeezed and displaced  state ($\psi(x)=\bra{x}S[r]D[\alpha]\ket{0}$) by solving the differential equation of annihilation operator $\hat{a}$ as follows
$$S[r]D[\alpha]\hat{a} D[-\alpha]S[-r]S[r]D[\alpha]\ket{0}=0. $$ 
For real squeezing ($r$) and complex displacement ($\alpha$), above equation in x-space using displacement identity $D(i Im[\alpha]) D(Re[\alpha])=D[\alpha]e^{iRe[\alpha]Im[\alpha]}$, turns into $$(\partial_{x}e^{-r}+x e^{r} -\sqrt{2}\alpha)\psi(x)=0,$$  leading to $$\psi(x)=\frac{1}{\sqrt[4]{(\pi e^{-2r})}}e^{-\frac{(x e^{r}-\sqrt{2}\alpha)^{2}}{2}-Im[\alpha]^{2}}e^{-iRe[\alpha]Im[\alpha]}.$$
Zeroth state  weight factor in squeezed displaced state $\psi(x)$ is given by
$$c_{0}(r,\alpha)=\langle0|S[r]D[\alpha]|0\rangle=\frac{e^{-\frac{\alpha ^2}{e^{2 r}+1}-i \alpha ^* \Im(\alpha )}}{\sqrt{\cosh (r)}}.$$
Number distribution and Wigner function are obtained through evaluation of cross term as follows
\begin{multline*}
    C_{n,m,k}^{\phi,r_{1},r_{2}}(\alpha,\beta,\delta,z)=\int_{-\infty}^{\infty}d^{2}\gamma \frac{\langle-\gamma|e^{i\phi a^{+}a}D[\delta]\widehat{O}|\gamma\rangle }{e^{-2(\gamma^* z-z^*\gamma)-2|z|^{2}}}
\\=\frac{\partial_{\phi}^{k}\partial_{s}^{n}\partial_{t}^{m}}{(i^{k}\sqrt{n!m!})}\left(\int_{-\infty}^{\infty}d^{2}\gamma\, e^{i \textnormal{Im}[e^{-i\phi}\gamma\delta^* -|\alpha|^{2}+(\bar{\alpha}-s)(s+\alpha^{*})]}\right.
\\\times\left.\left.  c_{0}(r_{1},\bar{\alpha})c_{0}^{*}(r_{2},\bar{\beta})\frac{e^{-e^{r_{2}}\gamma^{*}\beta+(\bar{\beta}^{*}-t)t}}{e^{-2(\gamma^* z-z^*\gamma)-2|z|^{2}}}e^{\frac{s^{2}+t^{2}}{2}}\right)\right|_{\tiny{\begin{array}{c} s=0\\t=0\end{array}}}\end{multline*}
where $z=x+ip,$ 
\begin{multline*}
    \widehat{O}=S[r_{1}]D[\alpha]|n\rangle\langle m|D[-\beta]S[-r_{2}]
\\=\left.\left(\partial_{s}^{n}\partial_{t}^{m}S[r_{1}]D[\alpha]|s\rangle\langle t|D[-\beta]S[-r_{2}]\frac{e^{\frac{s^2 +t^{2}}{2}}}{\sqrt{n!m!}}\right)\right|_{\tiny{\begin{array}{c} s=0\\t=0\end{array}}}\end{multline*} 
and $$\bar{\alpha}= e^{r_{1}}(\delta+e^{-i\phi}\gamma)+\alpha+s,\, \bar{\beta}=-e^{r_{2}}\gamma+\beta+t.$$
Above expression is normalized from cross-term  ($C_{n,m,k}^{\phi,r_{1},r_{2}}(\alpha,\beta,\delta,z)$) as 
$$\mathit{Nr}_{n,m,k}^{\phi,r_{1},r_{2}}(\alpha,\beta,\delta)=\int_{-\infty}^{\infty}d^{2}z\,\, C_{n,m,k}^{\phi,r_{1},r_{2}}(\alpha,\beta,\delta,z).$$ 

Using all these quantities above, we write the main text's symbols in the Table.\ref{tab:my_label1}.
\onecolumngrid

\begin{table}[H]
 \caption{\label{tab:my_label1} Used Symbols in The Main Text.}
   \centering
   \begin{ruledtabular}
    \begin{tabular}{cc}
  $I_{nm}(r_{1},r_{2},\alpha,\beta) = \left.\frac{\mathit{Nr}_{n,m,k}^{\phi,r_{1},r_{2}}(\alpha,\beta,0)}{\sqrt{\mathit{Nr}_{n,n,k}^{\phi,r_{1},r_{1}}(\alpha,\alpha,0)\mathit{Nr}_{m,m,k}^{\phi,r_{2},r_{2}}(\beta,\beta,0)}}\right|_{\tiny{\begin{array}{c}k=0\\ \phi=0\end{array}}}$    & $\bar{I}_{n}(r,\alpha,\beta) = \left.\frac{\mathit{Nr}_{n,n,k}^{\phi,r,r}(\alpha,\beta,0)}{\sqrt{\mathit{Nr}_{n,n,k}^{\phi,r,r}(\alpha,\alpha,0)\mathit{Nr}_{n,n,k}^{\phi,r,r}(\beta,\beta,0)}}\right|_{\tiny{\begin{array}{c} k=0\\\phi=0\end{array}}}$ \\ 
    $w_{nm}(r_{1},r_{2},\alpha,\beta) = \left.\frac{C_{n,m,k}^{\phi,r_{1},r_{2}}(\alpha,\beta,0,z)}{\sqrt{\mathit{Nr}_{n,n,k}^{\phi,r_{1},r_{1}}(\alpha,\alpha,0)\mathit{Nr}_{m,m,k}^{\phi,r_{2},r_{2}}(\beta,\beta,0)}}\right|_{\tiny{\begin{array}{c}k=0\\ \phi=0\end{array}}}$   & $w_{n}(r,\alpha,\beta) = \left.\frac{C_{n,n,k}^{\phi,r,r}(\alpha,\beta,0,z)}{\sqrt{\mathit{Nr}_{n,n,k}^{\phi,r,r}(\alpha,\alpha,0)\mathit{Nr}_{n,n,k}^{\phi,r,r}(\beta,\beta,0)}}\right|_{\tiny{\begin{array}{c} k=0\\\phi=0\end{array}}}$\\
 
      $ \mathit{Tr}\left[ D[\delta]\hat{O}\right]=
    \left.\frac{\mathit{Nr}_{n,m,k}^{\phi,r_{1},r_{2}}(\alpha,\beta,\delta)}{\sqrt{\mathit{Nr}_{n,n,k}^{\phi,r_{1},r_{1}}(\alpha,\alpha,0)\mathit{Nr}_{m,m,k}^{\phi,r_{2},r_{2}}(\beta,\beta,0)}}\right|_{\tiny{\begin{array}{c}k=0\\ \phi=0\end{array}}}$  &
    % \multicolumn{2}{c}{\hspace*{-1.3pt}   $\langle m|D[-\beta]S[-r_{2}]D[\delta]S[r_{1}]D[\alpha]|n\rangle=
    % \left.\frac{Nr_{n,m,k}^{\phi,r_{1},r_{2}}(\alpha,\beta,\delta)}{\sqrt{Nr_{n,n,k}^{\phi,r_{1},r_{1}}(\alpha,\alpha,0)Nr_{m,m,k}^{\phi,r_{2},r_{2}}(\beta,\beta,0)}}\right|_{\tiny{\begin{array}{c}k=0\\ \phi=0\end{array}}}$}\\
     $\mathit{Tr}\left[ \left(a^{+}a\right)^{k}\hat{O}\right]=
    \left.\frac{\mathit{Nr}_{n,m,k}^{\phi,r_{1},r_{2}}(\alpha,\beta,0)}{\sqrt{\mathit{Nr}_{n,n,k}^{\phi,r_{1},r_{1}}(\alpha,\alpha,0)\mathit{Nr}_{m,m,k}^{\phi,r_{2},r_{2}}(\beta,\beta,0)}}\right|_{\tiny{\begin{array}{c} \phi=0\end{array}}}$
\end{tabular}
\end{ruledtabular}
\end{table}

% \begin{multline*}
%  ,\end{multline*}
%     \begin{multline*}
%     \langle m|D[-\beta]S[-r_{2}](a^{+}a)^{k}S[r_{1}]D[\alpha]|n\rangle=\\
%     \left.\frac{Nr_{n,m,k}^{\phi,r_{1},r_{2}}(\alpha,\beta,0)}{\sqrt{Nr_{n,n,k}^{\phi,r_{1},r_{1}}(\alpha,\alpha,0)Nr_{m,m,k}^{\phi,r_{2},r_{2}}(\beta,\beta,0)}}\right|_{\tiny{\begin{array}{c} \phi=0\end{array}}}.\end{multline*}
   Two expressions in the last row being the cross-terms, are used to evaluate the overlap between perturbed and unperturbed state, average photon number, and number variance. 
    \twocolumngrid

\bibliography{biblio.bib}

\end{document}